\documentclass{jfm}
\usepackage{graphicx}
\usepackage{epstopdf}
\usepackage{float}
\usepackage{morefloats}
\usepackage{placeins}
\usepackage{xfrac}
\usepackage{pdflscape}
\usepackage{hhline}
\usepackage[table]{xcolor}
\usepackage{rotating}
\definecolor{light}{cmyk}{0,0,0,0.10}
\usepackage{tabularx}
\newcolumntype{Y}{>{\centering\arraybackslash}X}

\begin{document}

{ 
\newtheorem{lemma}{Lemma}
\newtheorem{corollary}{Corollary}

\shorttitle{Turbulent wall sheared thermal convection} 
\shortauthor{A. Blass \etal} 

\title{Flow organization and heat transfer in turbulent wall sheared thermal convection}

\author
{
	Alexander Blass\aff{1}
	\corresp{\email{a.blass@utwente.nl}},
	Xiaojue Zhu\aff{1,2},
	Roberto Verzicco\aff{3,1,4},\\
	Detlef Lohse\aff{1,5}
	\and 
	Richard J.A.M. Stevens\aff{1}
	\corresp{\email{r.j.a.m.stevens@utwente.nl}}	
}

\affiliation
{
	\aff{1}
	Physics of Fluids Group, Max Planck Center for Complex Fluid Dynamics, J. M. Burgers Center for Fluid Dynamics and MESA+ Research Institute, Department of Science and Technology, University of Twente, P.O. Box 217, 7500 AE Enschede, The Netherlands
	\aff{2}
	Center of Mathematical Sciences and Applications, School of Engineering and Applied Sciences, Harvard University, Cambridge, MA 02138, USA
	\aff{3}
	Dipartimento di Ingegneria Industriale, University of Rome "Tor Vergata". Via del Politecnico 1, Roma 00133, Italy
	\aff{4}
	Gran Sasso Science Institute - Viale F. Crispi, 7 67100 L'Aquila, Italy.
	\aff{5}
	Max Planck Institute for Dynamics and Self-Organization, Am Fassberg 17, 37077 G\"ottingen, Germany
}

\maketitle
}

\begin{abstract}
\noindent We perform direct numerical simulations of wall sheared Rayleigh-B\'enard (RB) convection for Rayleigh numbers up to $Ra=10^8$, Prandtl number unity, and wall shear Reynolds numbers up to $Re_w=10000$. Using the Monin-Obukhov length $L_{MO}$ we identify three different flow states, a buoyancy dominated regime ($L_{MO} \lesssim \lambda_{\theta}$; with $\lambda_{\theta}$ the thermal boundary layer thickness), a transitional regime ($0.5H \gtrsim L_{MO} \gtrsim \lambda_{\theta}$; with $H$ the height of the domain), and a shear dominated regime ($L_{MO} \gtrsim 0.5H$). In the buoyancy dominated regime the flow dynamics are similar to that of turbulent thermal convection. The transitional regime is characterized by rolls that are increasingly elongated with increasing shear. The flow in the shear dominated regime consists of very large-scale meandering rolls, similar to the ones found in conventional Couette flow. As a consequence of these different flow regimes, for fixed $Ra$ and with increasing shear, the heat transfer first decreases, due to the breakup of the thermal rolls, and then increases at the beginning of the shear dominated regime. For $L_{MO} \gtrsim 0.5H$ the Nusselt number $Nu$ effectively scales as $Nu \sim Ra^{\alpha}$, with $\alpha \ll 1/3$ while we find $\alpha \simeq 0.31$ in the buoyancy dominated regime. In the transitional regime the effective scaling exponent is $\alpha > 1/3$, but the temperature and velocity profiles in this regime are not logarithmic yet, thus indicating transient dynamics and not the ultimate regime of thermal convection.
\end{abstract} 

\section{Introduction} \label{Introduction}
\noindent \textcolor{black}{Rayleigh-B\'enard (RB) convection, i.e. the flow in a box heated from below and cooled from above, is one of the paradigmatic fluid dynamical systems \citep{ahl09,loh10,chi12,xia13}. The dynamics of RB convection are controlled by the Rayleigh number
	\begin{equation} 
	Ra=\beta gH^3\Delta /(\kappa\nu),
	\label{eqn:Ra} \\
	\end{equation}	
which is the non-dimensional temperature difference between the horizontal plates, and the Prandtl number 
	\begin{equation} 
	Pr=\nu/\kappa,
	\label{eqn:Pr} \\
	\end{equation}	
which is the ratio of momentum and thermal diffusivities. In equations (\ref{eqn:Ra}) and (\ref{eqn:Pr}), $H$ is the non-dimensional distance between the plates, $\beta$ the thermal expansion coefficient of the fluid, $g$ the gravitational acceleration, $\Delta$ the temperature difference between top and bottom plate, and $\kappa$ and $\nu$ the thermal and kinetic diffusivities, respectively. The crucial additional control parameter, whose effect we will analyse in this paper, is the wall shear Reynolds number 
	\begin{equation} 
	Re_w=H u_w/\nu,
	\label{eqn:Re_w} \\
	\end{equation}
with $u_w$ the velocity of the wall. The ratio between buoyancy and shear driving can be expressed as bulk Richardson number 	
	\begin{equation} 
	Ri=Ra/(Re^2_wPr),
	\label{eqn:Ri} \\
	\end{equation}
which can be seen as alternative control parameter for either $Ra$ or $Re_w$.}

\textcolor{black}{Important responses of the system are the Nusselt number 
\begin{equation} 
	Nu= QH/(\kappa \Delta),
	\label{eqn:Nu} \\
	\end{equation}	
which is the dimensionless vertical heat flux, the friction Reynolds number 
	\begin{equation} 
	Re_\tau=Hu_\tau/ \nu,
	\label{eqn:Re_t} \\
	\end{equation}
and the skin friction coefficient	
	\begin{equation} 
	C_f= 2\tau_w / \rho u_w ^2.
	\label{eqn:C_f} \\
	\end{equation}	
Here $Q=\overline{w' \theta '} - \kappa \partial T /\partial z$ is the constant vertical heat flux, with $w'$ and $\theta '$ the fluctuations for wall-normal velocity and temperature, respectively, and $u_\tau = \sqrt{\tau _w / \rho}$ the friction velocity, with $\tau _w$ the mean wall shear stress and $\rho$ the density of the fluid.}

\begin{figure}
	\includegraphics[width=\textwidth]{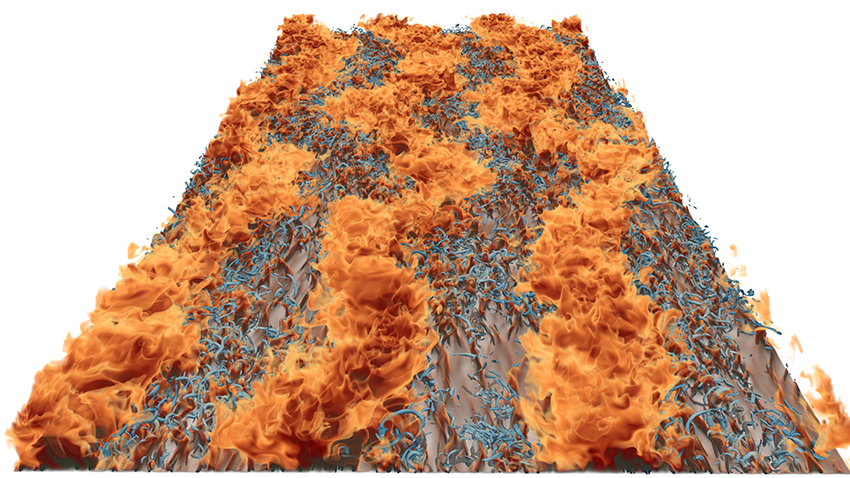}%
	\caption{\label{fig:3Dpic} \textcolor{black}{Volume rendering} of the thermal structures rising from the heated plate in a simulation with $Ra=2.2\times10^6$ and $Re_w=8000$. The plate dimensions are $9 \pi H \times 4 \pi H$, in streamwise and spanwise direction, respectively, where $H$ is the distance between the plates. The red colors show hot thermal structures emerging from the hot plate while the blue structures show vorticity formations in the flow. \textcolor{black}{For further details of the flow visualization, please see \cite{fav19}.}}
\end{figure}

For strong enough thermal driving, i.e. high enough $Ra$, the flow in the bulk region becomes fully turbulent. For even stronger thermal driving, beyond some critical $Ra$ number $Ra_c$, also the boundary layers become turbulent and the system reaches the regime of so-called ultimate convection \citep{kra62,gro00,gro01,gro11}. This ultimate regime sets in when the shear Reynolds number at the boundary layers is sufficiently high so that the boundary layer becomes turbulent, leading to a strong increase in the heat transport, \textcolor{black}{quantified by the Nusselt number}.

\cite{ahl12b} found that the transition to the ultimate regime sets in around $Ra_c \sim \mathcal{O}(10^{14})$. While in the classical regime one generally finds $Nu \sim Ra^{0.31}$, in the ultimate regime $Nu \sim Ra^{0.38}$, in agreement with theoretical predictions \citep{gro11}. 

The transition to the ultimate regime has also been observed in direct numerical simulations (DNS) of two-dimensional RB convection \citep{zhu18c}. In Taylor-Couette flow, which is a very analogous system, experiments and DNS have observed the ultimate regime as well \citep{gro16}. \textcolor{black}{Such scaling has also been observed in experiments with vertical pipes \citep{gib06,cho09,paw16}.} However, so far the ultimate regime has not yet been achieved in DNS of three-dimensional RB flows \citep{ste10,ste11} as the required computational time to achieve this is still out of reach. In an attempt to trigger the transition to the ultimate regime, here we add a Couette type shearing to the RB system to increase the shear Reynolds number of the boundary layers.

In Couette flow the top and bottom walls move in opposite directions \citep{thu00,bar05,tuc11} \textcolor{black}{with constant $u_w$} and just as in other examples of wall bounded turbulence \citep{jim18,smi11,smi13} the flow is dominated by elongated streaks, which have been observed in experiments \citep{kit08} and DNS \citep{lee91,tsu06}, even at relatively low shear Reynolds numbers \citep{cha17}. \cite{pir11,pir14} and \cite{orl15} showed that these streaks in Couette flow have much longer characteristic length scales than in Poiseuille flow, where the flow is forced by a uniform pressure gradient rather than by wall shear. \cite{raw15} showed that these large-scale flow structures even survive when the small-scale structures are artificially surpressed. Recently, \cite{lee18} found that the streak length increases with increasing shear Reynolds number and that some correlation in the streamwise direction remains visible up to a length of almost 80 times the distance between the plates.

\begin{figure}
	\centering
	\includegraphics[width=\textwidth]{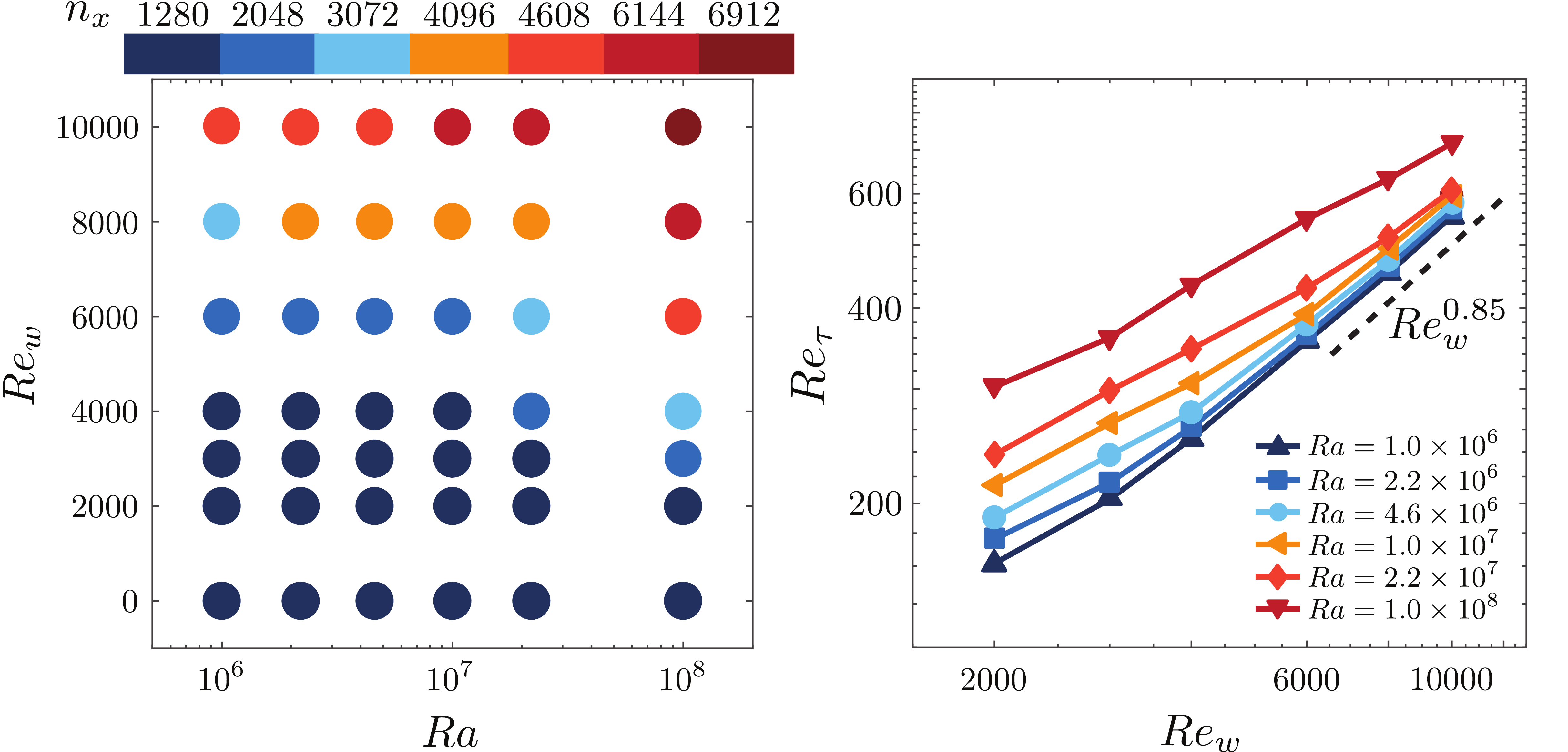}%
	\caption{\label{fig:status} (a) Streamwise ($n_x$) resolution used in the simulations as function of $Ra$ and $Re_w$, see table \ref{tab:overviewcases} for details. (b) $Re_\tau$ versus $Re_w$ for the simulations. In agreement with \cite{pir14} and \cite{avs14}, for large $Re_w$ we find $Re_{\tau} \sim Re_w^{0.85}$.}
\end{figure}

Investigating the interaction between buoyancy and shear effects is also very important to better understand oceanic and atmospheric flows \citep{dea72,moe84,kha98}. For example early experiments on sheared thermal convection by \cite{ing66} and \cite{sol90} showed the appearance of large-scale structures. \cite{fuk85} showed that in channel flow unstable stratification increases the longitudinal velocity fluctuations close to the wall, while in the bulk region the temperature fluctuations are drastically lowered. 

Furthermore, recent experiments by \cite{she19} investigated the plume spacing in sheared convection and found a scaling law which connects the mean spacing of the plumes with $Re_w$, $Ra$, and $Pr$ of the flow.

Early simulations of sheared convection were performed by \cite{hat86} and \cite{dom88} for $Ra\lesssim\mathcal{O}(10^5)$. \cite{dom88} found that in Couette-RB flow the addition of shear at low $Ra$ initially increases the heat transport. However, for $Ra\gtrsim150.000$ the heat transport decreases as the added shear breaks up the large-scale structures. More recently, \cite{sca14,sca15} showed that also in Poiseuille-RB the heat transfer first decreases when the applied pressure gradient is increased. The reason is that for intermediate forcing the longitudinal wind disturbs the thermal plumes which therefore lose their coherence. Only with a strong enough pressure gradient, a heat transfer enhancement is found. \textcolor{black}{\cite{sca14} in particular find how $Nu$ depends on $Ra$, $Pr$, and $Re_\tau$.}

\textcolor{black}{Forced convection in turbulent Couette flow has been investigated theoretically \citep{cho04}, numerically \citep{liu03,deb04}, and experimentally \citep{le06} and shows similar behavior as the high-shear cases of sheared convection.}

The Richardson number quantifies the ratio between the buoyancy and shear forces in Couette-RB and Poiseuille-RB based on the applied temperature difference and wall shear Reynolds number. \textcolor{black}{Another way to quantify the ratio between buoyancy and shear forces is to determine the Monin-Obukhov length \citep{mon54,obu71}
\begin{equation} 
L_{MO}/H=u^3_\tau /(\overline{w' \theta '} \beta g H),
\label{eqn:L} \\
\end{equation}
which indicates up to which distance from the wall the flow is dominated by shear, based on the observed flow properties. Note that $L_{MO}/H$ is a response parameter, in contrast to $Ri$, which is a control parameter. \cite{pir17} found that the Monin-Obukhov length scales as $L_{MO}/H \approx 0.15/ Ri^{0.85}$ for channel flow with unstable stratification.} 

In this study we investigate the effect of an additional Couette type shearing on the heat transfer in RB convection in an attempt to trigger the boundary layers to become fully turbulent and hence observe the transition to the ultimate regime. Figure \ref{fig:3Dpic} shows a flow visualization of the temperature field obtained from one of our simulations, which reveals large-scale meandering streaks that are formed near the hot plate. We performed simulations over a wide parameter range, spanning $10^6 \leq Ra \leq 10^8$ and $0 \leq Re_w \leq 10^4$, while $Pr=1$ has been used in all cases, see figure \ref{fig:status}a. In spite of the very strong forcing for the largest $Ra$ and $Re_w$, we did not yet achieve ultimate turbulence in this study. We were limited by our own requirement of using large domain sizes as recommended by \cite{pir17} to ensure convergence of the main flow properties.

\begin{table} 
	\vspace{-0.3cm}	
	\begin{center}
		\begin{tabularx}{\textwidth}{c@{}Y@{}Y@{}Y@{}Y@{}Y@{}Y@{}Y@{}Y@{}Y@{}Y@{}YYY} 
			$Ra$ & $Pr$ & $Re_w$ & & $N_x$ & $N_y$ & $N_z$ & & $Re_{\tau}$ & $Ri$ & & $L_{MO}/H$ & $Nu$ & $C_f/10^{-3}$ \\[1pt]
			
			\hline \hline \\[-5.5pt]
			
			$ 1.0\times10^{8} $ & 1 & 10000 & & 1024 & 512 & 384 & & 708.0 & 1.000 & & 0.138 & 25.66 & 10.03 \\
			$ 1.0\times10^{8} $ & 1 & 10000 & & 1296 & 648 & 384 & & 703.6 & 1.000 & & 0.137 & 25.37 & 9.902 \\
			$ 1.0\times10^{8} $ & 1 & 10000 & & 1536 & 768 & 384 & & 700.6 & 1.000 & & 0.137 & 25.19 & 9.818 \\
			$ 1.0\times10^{8} $ & 1 & 10000 & & 1728 & 864 & 384 & & 700.0 & 1.000 & & 0.136 & 25.15 & 9.805 \\ [1pt]
			
			\hline \hline \\[-6.5pt]
			
		\end{tabularx}
		\caption{Simulation parameters for the grid study, which is performed in a box of $2 \pi H \times \pi H \times H$. The columns from left to right indicate the input and output parameters and the resolution in streamwise, spanwise, and wall-normal direction $(N_x,N_y,N_z)$. }
		{\label{tab:gridcases}}
	\end{center}
\end{table}

\begin{figure}
	\centering
	\includegraphics[width=\textwidth]{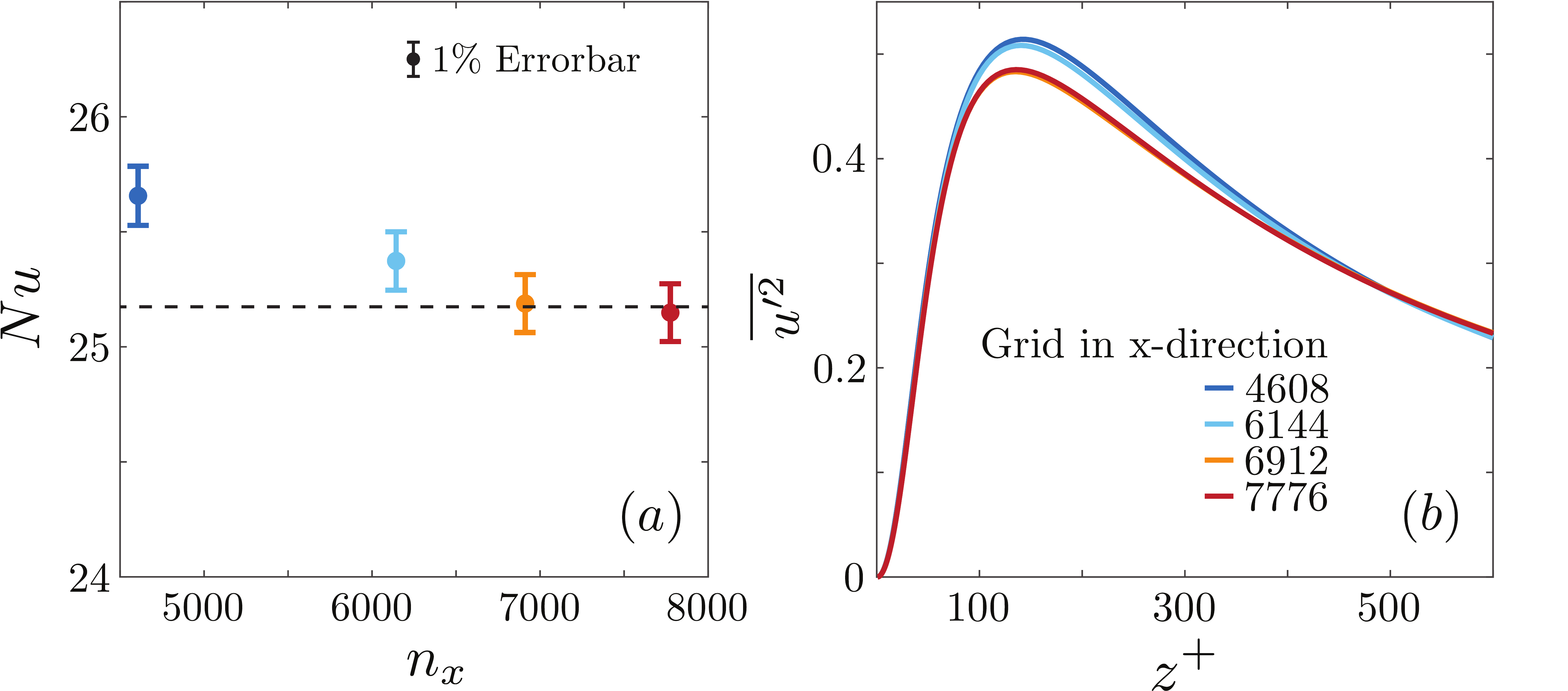}%
	\caption{\label{fig:gridstudy} (a) $Nu$ and (b) the streamwise velocity fluctuations for simulations at $Ra=10^8$ and $Re_w =10000$ performed in a box of $2 \pi H \times \pi H \times H$ in streamwise, spanwise, and vertical direction, respectively, on different grids. The displayed resolutions indicate the extrapolated streamwise resolutions that correspond to the full $9\pi H \times 4\pi H \times H$ box, see table \ref{tab:gridcases} for details. Note that the simulation results are converged for the grid resolution used in this study.}
\end{figure}

The remainder of this manuscript is organized as follows. In \S \ref{NumSim} we present the simulation method. We discuss the heat transfer and skin friction measurements in \S \ref{Nu} and \S \ref{skinfriction}, respectively. A discussion of the identified flow regimes is given in \S \ref{local}. The concluding remarks follow in \S \ref{conclusion}.

\begin{table} 
	\vspace{-0.3cm}	
	\begin{center}
		\begin{tabularx}{\textwidth}{c@{}Y@{}Y@{}Y@{}Y@{}Y@{}Y@{}Y@{}Y@{}Y@{}Y@{}YYY} 
			$Ra$ & $Pr$ & $Re_w$ & & $N_x$ & $N_y$ & $N_z$ & & $Re_{\tau}$  &   $Ri$   & & $L_{MO}/H$ & $Nu$  & $C_f/10^{-3}$ \\[1pt]
			
			\hline \hline \\[-6.5pt]
			
			$ 0			      $ & 1 &  2000 & & 1280 & 1024 & 256 & & 134.0 &        0 & &     0 &     0 & 8.971 \\
			$ 0			      $ & 1 &  3000 & & 1280 & 1024 & 256 & & 188.7 &        0 & &     0 &     0 & 7.917 \\
			$ 0			      $ & 1 &  4000 & & 1280 & 1024 & 256 & & 241.8 &        0 & &     0 &     0 & 7.314 \\
			$ 0 			  $ & 1 &  6000 & & 1280 & 1024 & 256 & & 346.0 &        0 & &     0 &     0 & 6.651 \\
			$ 0 			  $ & 1 &  8000 & & 2048 & 1280 & 256 & & 400.8 &        0 & &     0 &     0 & 5.020 \\
			$ 0				  $ & 1 & 10000 & & 3072 & 1536 & 256 & & 468.0 &        0 & &     0 &     0 & 4.382 \\ [1pt]
			
			\hline \\[-6.5pt]		
			
			$ 1.0\times10^{6} $ & 1 &     0 & & 1280 & 1024 & 256 & &     0 & $\infty$ & & 0 	 & 8.343 & $\infty$ \\
			$ 1.0\times10^{6} $ & 1 &  2000 & & 1280 & 1024 & 256 & & 164.2 &    0.250 & & 0.552 & 6.557 & 13.49 \\
			$ 1.0\times10^{6} $ & 1 &  3000 & & 1280 & 1024 & 256 & & 206.2 &    0.111 & & 1.180 & 6.869 & 9.449 \\
			$ 1.0\times10^{6} $ & 1 &  4000 & & 1280 & 1024 & 256 & & 255.7 &    0.063 & & 1.983 & 7.891 & 8.173 \\
			$ 1.0\times10^{6} $ & 1 &  6000 & & 2048 & 1280 & 256 & & 360.7 &    0.028 & & 4.258 & 10.52 & 7.231 \\
			$ 1.0\times10^{6} $ & 1 &  8000 & & 3072 & 1536 & 256 & & 459.3 &    0.016 & & 7.211 & 12.82 & 6.592 \\
			$ 1.0\times10^{6} $ & 1 & 10000 & & 4608 & 2304 & 320 & & 563.0 &    0.010 & & 11.00 & 15.49 & 6.340 \\ [1pt]
			
			\hline \\[-6.5pt]
			
			$ 2.2\times10^{6} $ & 1 &     0 & & 1280 & 1024 & 256 & &     0 & $\infty$ & &    0  & 10.40 & $\infty$ \\
			$ 2.2\times10^{6} $ & 1 &  2000 & & 1280 & 1024 & 256 & & 179.0 &    0.550 & & 0.306 & 7.866 & 16.03 \\
			$ 2.2\times10^{6} $ & 1 &  3000 & & 1280 & 1024 & 256 & & 218.7 &    0.244 & & 0.582 & 7.788 & 10.63 \\
			$ 2.2\times10^{6} $ & 1 &  4000 & & 1280 & 1024 & 256 & & 266.1 &    0.138 & & 0.953 & 8.568 & 8.848 \\
			$ 2.2\times10^{6} $ & 1 &  6000 & & 2048 & 1280 & 256 & & 368.7 &    0.061 & & 2.021 & 10.96 & 7.551 \\
			$ 2.2\times10^{6} $ & 1 &  8000 & & 4096 & 2048 & 256 & & 470.0 &    0.034 & & 3.350 & 13.44 & 6.904 \\
			$ 2.2\times10^{6} $ & 1 & 10000 & & 4608 & 2304 & 320 & & 575.3 &    0.022 & & 5.123 & 16.12 & 6.619 \\ [1pt]
			
			\hline \\[-6.5pt]
			
			$ 4.6\times10^{6} $ & 1 &     0 & & 1280 & 1024 & 256 & &     0 & $\infty$ & &    0  & 12.83 & $\infty$ \\
			$ 4.6\times10^{6} $ & 1 &  2000 & & 1280 & 1024 & 256 & & 193.2 &    1.150 & & 0.139 & 9.353 & 18.66 \\
			$ 4.6\times10^{6} $ & 1 &  3000 & & 1280 & 1024 & 256 & & 241.0 &    0.511 & & 0.316 & 9.502 & 12.91 \\
			$ 4.6\times10^{6} $ & 1 &  4000 & & 1280 & 1024 & 256 & & 280.4 &    0.288 & & 0.466 & 9.626 & 9.829 \\
			$ 4.6\times10^{6} $ & 1 &  6000 & & 2048 & 1280 & 256 & & 383.1 &    0.128 & & 0.982 & 11.88 & 8.154 \\
			$ 4.6\times10^{6} $ & 1 &  8000 & & 4096 & 2048 & 256 & & 481.5 &    0.072 & & 1.644 & 14.08 & 7.244 \\
			$ 4.6\times10^{6} $ & 1 & 10000 & & 4608 & 2304 & 320 & & 587.7 &    0.046 & & 2.512 & 16.76 & 6.910 \\ [1pt]
			
			\hline \\[-6.5pt]
			
			$ 1.0\times10^{7} $ & 1 &     0 & & 1280 & 1024 & 256 & &     0 & $\infty$ & &     0 & 16.18 & $\infty$ \\
			$ 1.0\times10^{7} $ & 1 &  2000 & & 1280 & 1024 & 256 & & 216.5 &    2.500 & & 0.075 & 12.41 & 23.45 \\
			$ 1.0\times10^{7} $ & 1 &  3000 & & 1280 & 1024 & 256 & & 269.9 &    1.111 & & 0.156 & 12.02 & 16.19 \\
			$ 1.0\times10^{7} $ & 1 &  4000 & & 1280 & 1024 & 256 & & 310.7 &    0.625 & & 0.233 & 11.78 & 12.07 \\
			$ 1.0\times10^{7} $ & 1 &  6000 & & 2048 & 1280 & 256 & & 397.3 &    0.278 & & 0.475 & 12.85 & 8.771 \\
			$ 1.0\times10^{7} $ & 1 &  8000 & & 4096 & 2048 & 256 & & 501.1 &    0.156 & & 0.785 & 15.30 & 7.848 \\
			$ 1.0\times10^{7} $ & 1 & 10000 & & 6144 & 3072 & 320 & & 604.3 &    0.100 & & 1.218 & 17.85 & 7.305 \\ [1pt]			
			
			\hline \\[-6.5pt]
			
			$ 2.2\times10^{7} $ & 1 &     0 & & 1280 & 1024 & 256 & &     0 & $\infty$ & &     0 & 20.92 & $\infty$ \\
			$ 2.2\times10^{7} $ & 1 &  2000 & & 1280 & 1024 & 256 & & 241.4 &    5.500 & & 0.036 & 16.97 & 29.15 \\
			$ 2.2\times10^{7} $ & 1 &  3000 & & 1280 & 1024 & 256 & & 302.9 &    2.445 & & 0.078 & 16.13 & 20.40 \\
			$ 2.2\times10^{7} $ & 1 &  4000 & & 2048 & 1280 & 256 & & 351.1 &    1.375 & & 0.118 & 15.90 & 15.42 \\
			$ 2.2\times10^{7} $ & 1 &  6000 & & 3072 & 1536 & 256 & & 435.8 &    0.611 & & 0.243 & 15.43 & 10.56 \\
			$ 2.2\times10^{7} $ & 1 &  8000 & & 4096 & 2048 & 256 & & 522.0 &    0.344 & & 0.374 & 16.52 & 8.517 \\
			$ 2.2\times10^{7} $ & 1 & 10000 & & 6144 & 3072 & 320 & & 613.9 &    0.220 & & 0.644 & 18.01 & 7.541 \\ [1pt]
			
			\hline \\[-6.5pt]
			
			$ 1.0\times10^{8} $ & 1 &     0 & & 1280 & 1024 & 256 & &     0 & $\infty$ & &     0 & 36.52 & $\infty$ \\
			$ 1.0\times10^{8} $ & 1 &  2000 & & 1280 & 1024 & 256 & & 307.2 &    25.00 & & 0.009 & 35.06 & 47.20 \\
			$ 1.0\times10^{8} $ & 1 &  3000 & & 2048 & 1280 & 256 & & 365.5 &    11.11 & & 0.014 & 29.18 & 29.68 \\
			$ 1.0\times10^{8} $ & 1 &  4000 & & 3072 & 1536 & 256 & & 440.6 &    6.250 & & 0.030 & 27.23 & 24.27 \\
			$ 1.0\times10^{8} $ & 1 &  6000 & & 4608 & 2304 & 320 & & 557.0 &    2.778 & & 0.060 & 26.27 & 17.24 \\
			$ 1.0\times10^{8} $ & 1 &  8000 & & 6144 & 3072 & 320 & & 660.6 &    1.563 & & 0.097 & 26.00 & 13.64 \\
			$ 1.0\times10^{8} $ & 1 & 10000 & & 6912 & 3456 & 384 & & 740.2 &    1.000 & & 0.160 & 25.21 & 10.96 \\ [1pt]
			
			\hline \hline \\[-6.5pt]
			
		\end{tabularx}
		\caption{Main simulations considered in this work.} 
		{\label{tab:overviewcases}}
	\end{center}
\end{table}

\section{Simulation details} \label{NumSim}
\noindent We numerically solve the three-dimensional incompressible Navier-Stokes equations within the Boussinesq approximation, which in non-dimensional form read: 
\begin{equation} 
\frac{\partial \boldsymbol{u}}{\partial t} + \boldsymbol{u} \bcdot \bnabla \boldsymbol{u} =-\bnabla P + \left(\frac{Pr}{Ra} \right)^{1/2} \nabla^2\boldsymbol{u}+\theta \hat{z}, \mbox{~~~~} \bnabla \bcdot \boldsymbol{u} =0,
\label{eqn:Navier}
\end{equation}
\begin{equation} 
\frac{\partial \theta}{\partial t} + \boldsymbol{u} \bcdot \bnabla \theta = \frac{1}{(Pr Ra)^{1/2}} \nabla ^2 \theta,
\label{eqn:temp} \\
\end{equation}
\textcolor{black}{with $\boldsymbol{u}$ the velocity non-dimensionalized by the free-fall velocity $\sqrt{g \beta \Delta H}$, $t$ the dimensionless time normalized by $\sqrt{H/(g \beta \Delta)}$, $\theta$ the dimensionless temperature normalized by the temperature difference between the plates $\Delta$, and $P$ the pressure normalized by $g \beta \Delta /H$.} 

To solve equations (\ref{eqn:Navier}) - (\ref{eqn:temp}) we employ the second-order finite difference code AFiD \citep{poe15c}, which has been validated many times against other numerical and experimental results \citep{ver96,ver97,ver03,ste10,ste11,ost14d,koo18}. The code uses periodic boundary conditions with uniform mesh spacing in the horizontal directions and supports a non-uniform grid distribution in the wall-normal direction. For this study we used the GPU version of the code \citep{zhu18b} to allow efficient execution of many large-scale simulations. The Couette flow forcing is realized by moving both walls in opposite directions with speed $u_w$ and the results for the pure Couette flow case match excellently with the results by \cite{pir14}. For example, figure \ref{fig:status}b shows that for Couette flow $Re_{\tau} \sim Re_w^{0.85}$, which agrees very well with the Couette data of \cite{pir14} and \cite{avs14}. 

All simulations in this study were performed in a large $9\pi H \times 4\pi H \times H$ box, in streamwise, spanwise and wall-normal direction \citep{tsu06,pir14}, which is required to capture the large-scale structures formed in the Couette \citep{pir14,avs14,lee18}. \textcolor{black}{We adopted the grid distribution used by \cite{pir14,pir17}, which is based on the resolution requirements for pure buoyant flow \citep{shi10} and pure channel flow \citep{ber14}, which is very similar to our flow configuration. As initial condition for our code we use previous flow fields and we make sure that all simulations are statistically stable before extracting data to ensure an independence on the initial conditions. We performed additional simulations with varying grid resolutions as additional grid refinement check for the $Ra=10^8$ and $Re_w =10000$ case, i.e. the most challenging simulation of this study.} To keep this resolution test manageable it is performed in a smaller $2\pi H \times \pi H \times H$ domain, see table \ref{tab:gridcases}. Figure \ref{fig:gridstudy} confirms that the simulations are fully resolved for the chosen resolution. As a further validation, we evaluate the Couette data from \cite{pir14} in \S \ref{skinfriction}, which collapses very well with our data. Table \ref{tab:overviewcases} shows the simulation parameters for the main cases presented in this study.

\begin{figure}
	\centering
	\includegraphics[width=0.7\textwidth]{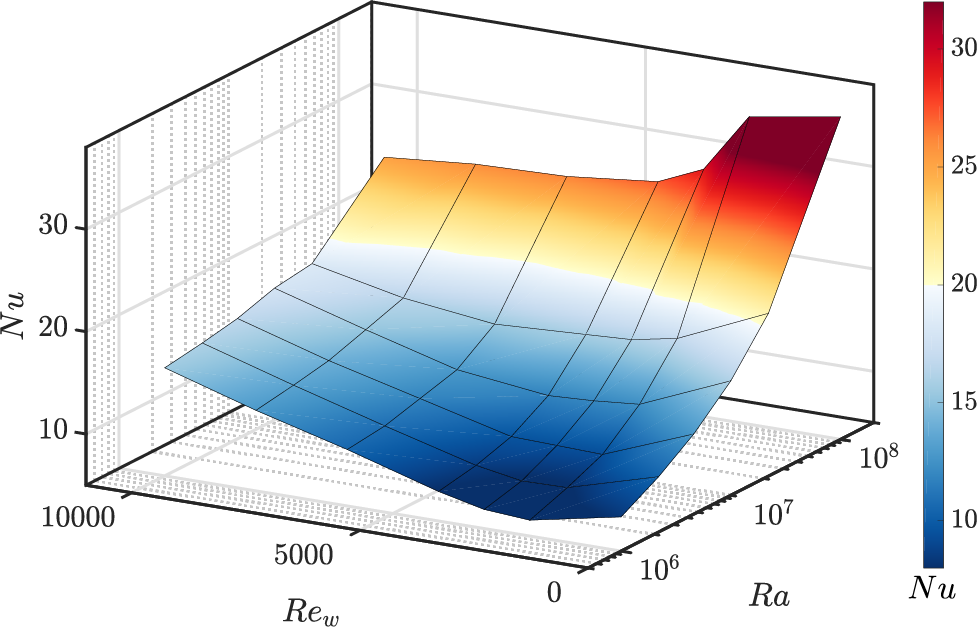}%
	\caption{\label{fig:nusselt3d} $Nu$ as a function of $Ra$ and $Re_w$ in Couette-RB flow.}
\end{figure}

\section{Global flow characteristics} \label{results}

\subsection{Effective scaling of the Nusselt number} \label{Nu}
\begin{figure}
	\includegraphics[width=\textwidth]{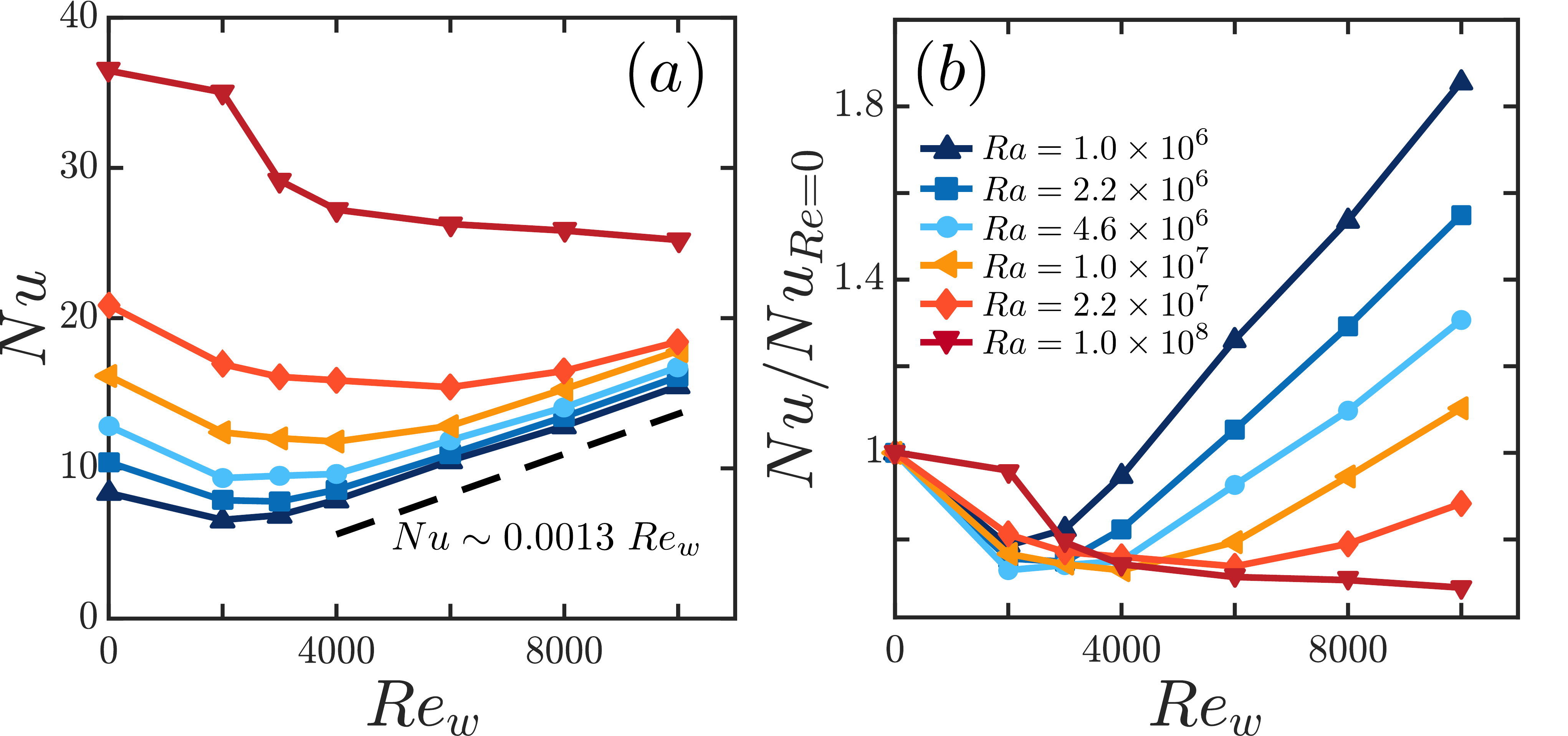}%
	\caption{\label{fig:nusselt} (a) $Nu$ and (b) $Nu$ normalized by the RB value $Nu_{(Re_w=0)}$ as a function of $Re_w$.}
	\centering
\end{figure}

\noindent Figure \ref{fig:nusselt3d} shows that the heat transfer increases with increasing $Ra$ and $Re_w$ and that for given $Ra$ number a minimum heat transfer is obtained at some intermediate $Re_w$. Figure \ref{fig:nusselt}a shows the corresponding cross sections for constant $Ra$ which clearly reveal that the location of the minimum heat transfer at constant $Ra$ shifts towards higher $Re_w$ with increasing $Ra$. \textcolor{black}{For high enough $Re_w$, the behavior of $Nu$ converges towards $Nu \sim 0.0013 Re_w$. In panel \ref{fig:nusselt}b, where $Nu$ is normalized by the RB value for the respective $Ra$, we can see very clearly that for low $Re_w$ and with increasing $Ra$ the thermal plumes become stronger and therefore harder to disturb by the applied shear. For $Ra=10^8$ the decrease in $Nu$ at $Re_w=2000$ is only $\sim 4 \%$ while the data for other $Ra$ show percentages in the high twenties. A more exact analysis would need more detailed datapoints for low $Re_w$.}

The results indicate that this mechanism is influenced by the ratio of the buoyancy and shear forces. Therefore the bulk Richardson number $Ri$ or the above defined Monin-Obukhov length $L_{MO}$, which take the ratio of these forces into account, are natural control and response parameters to identify the different flow regimes. \textcolor{black}{Although the Monin-Obukhov theory itself is only valid for shear dominated flow, which does not necessarily exist in parts of our flow simulations, we use this parameter as an objective criterion to distinct between buoyancy and shear driven flow.} As $L_{MO}$ can be compared to other important length scales in the flow, we decide to use it for this purpose. From the data in table \ref{tab:overviewcases}, we find $L_{MO}/H \approx 0.16 /Ri^{0.91}$ (see appendix \ref{app:Ri_L}). In figure \ref{fig:monin}a the Monin-Obukhov length is compared to the thermal boundary layer thickness $\lambda_{\theta}$ and the arbitrary threshold $0.5H$ is reported for later discussion. Since $L_{MO}/H$ is the fraction of the domain in which the \textcolor{black}{shear} forcing is dominant in the flow, $L_{MO} \geq 0.5H$ is the threshold from which on the flow is completely shear dominated since the wall generated shear affects at least half of the fluid layer thickness. This allows us to define three different flow regimes, namely a buoyancy dominated regime ($L_{MO} \lesssim \lambda_{\theta}$), a transitional regime ($0.5H \gtrsim L_{MO} \gtrsim \lambda_{\theta}$), and a shear dominated regime ($L_{MO} \gtrsim 0.5H$). A similar behavior has also been observed in convective boundary layers, where \cite{sal16} find a cell dominated regime for $z_i/L_{MO}>20$, where $z_i$ is the convective boundary layer thickness, a cell and roll dominated regime as transitional state, and a roll dominated regime for $z_i/L_{MO}<5$.

\begin{figure}
	\centering
	\includegraphics[width=\textwidth]{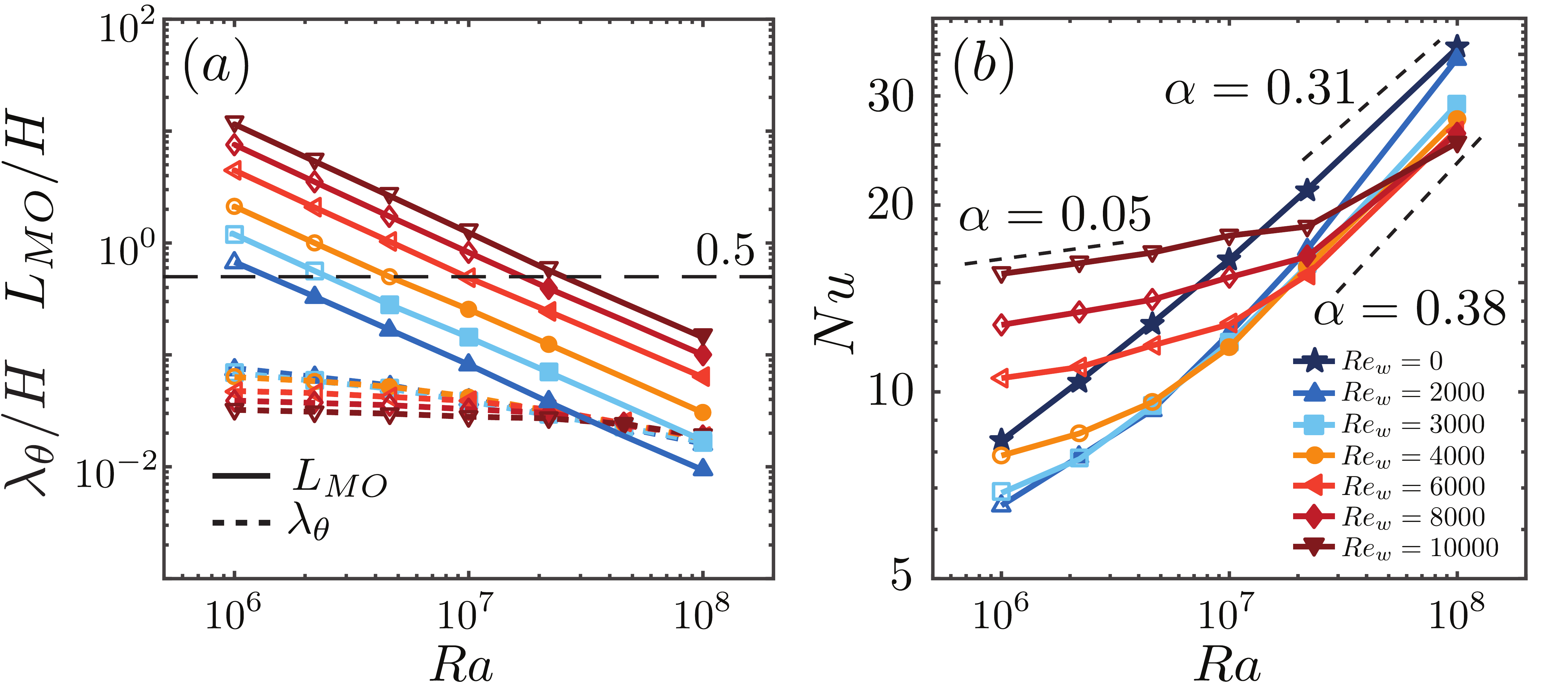}%
	\caption{\label{fig:monin} (a) The Monin-Obukhov length $L_{MO}$ as function of $Ra$ for different $Re_w$. $L_{MO}$ is compared to the thermal boundary layer thickness $\lambda _\theta$ and $0.5H$ to define the flow regime (temperature dominated, transition, shear dominant) of each simulations, see details in the text. (b) $Nu$ as function of $Ra$. The numbers indicate the scaling exponent $\alpha$ in $Nu \sim Ra ^\alpha$.}
\end{figure}

 Figure \ref{fig:monin}b shows that the heat transfer in the buoyancy dominated regime scales as $Nu \simeq Ra^{0.31}$, as also found for classical RB convection ($Re_w=0$, \cite{ahl09}). For the shear dominated regime we find that the effective scaling exponent $\alpha$ in $Nu \sim Ra^\alpha$ is $\alpha \ll 1/3$ and in the transitional regime we find $\alpha>1/3$. An effective scaling exponent larger than $1/3$ is one of the characteristics of the ultimate regime. It should occur when the boundary layers have transitioned to the turbulent state, which is indicated by their logarithmic profiles. Our analysis in \S 4 will show that this is not yet the case in this transitional regime. Instead, for intermediate shear, the heat transfer is decreased with respect to the RB case. The locally larger effective scaling exponent simply is a consequence of the fact that with increasing $Ra$ the heat transfer, \textcolor{black}{which was decreased at intermediate shear,} must again converge to the RB case.

\subsection{Skin Friction} \label{skinfriction}

\begin{figure}
	\centering
	\includegraphics[width=\textwidth]{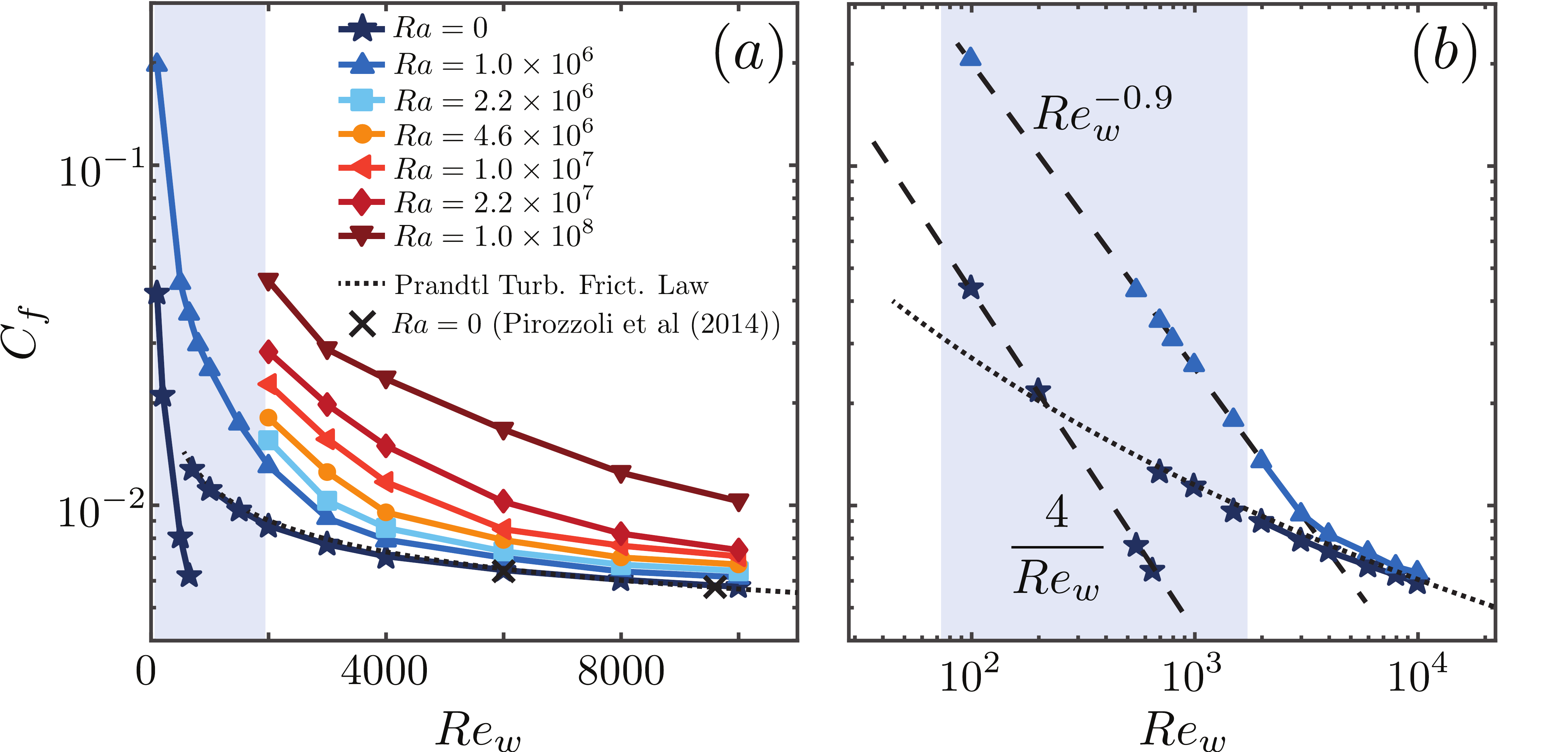}%
	\caption{\label{fig:cf} (a) Skin friction coefficient $C_f$ as a function of $Re_w$. (b) Zoom-in of the gray area shown in panel (a), now on a log-scale, showing the data for pure Couette flow ($Ra=0$, stars) and $Ra=10^6$. Note that $C_f(Ra=0)$ follows the expected laminar result (- - -) until $Re_w=650-700$ and then jumps to the turbulent curve ($ \cdot \cdot \cdot $). For Couette-RB, i.e. the uppointing triangle, no jump is observed.}
\end{figure}
\noindent In figure \ref{fig:cf} we compare the measured skin friction coefficient for different $Re_w$ and $Ra$ with Prandtl's turbulent friction law \citep{sch00}:
\begin{equation} 
\sqrt{\frac{2}{C_f}}= \frac{1}{\mathcal{K}} \log \left(Re_w \sqrt{\frac{C_f}{2}} \right)+C.
\label{eqn:cfprandtl}
\end{equation}

\begin{figure}
	\vspace{1cm}
	\centering
	\rotatebox{90}{
		\includegraphics[width=0.8\textheight]{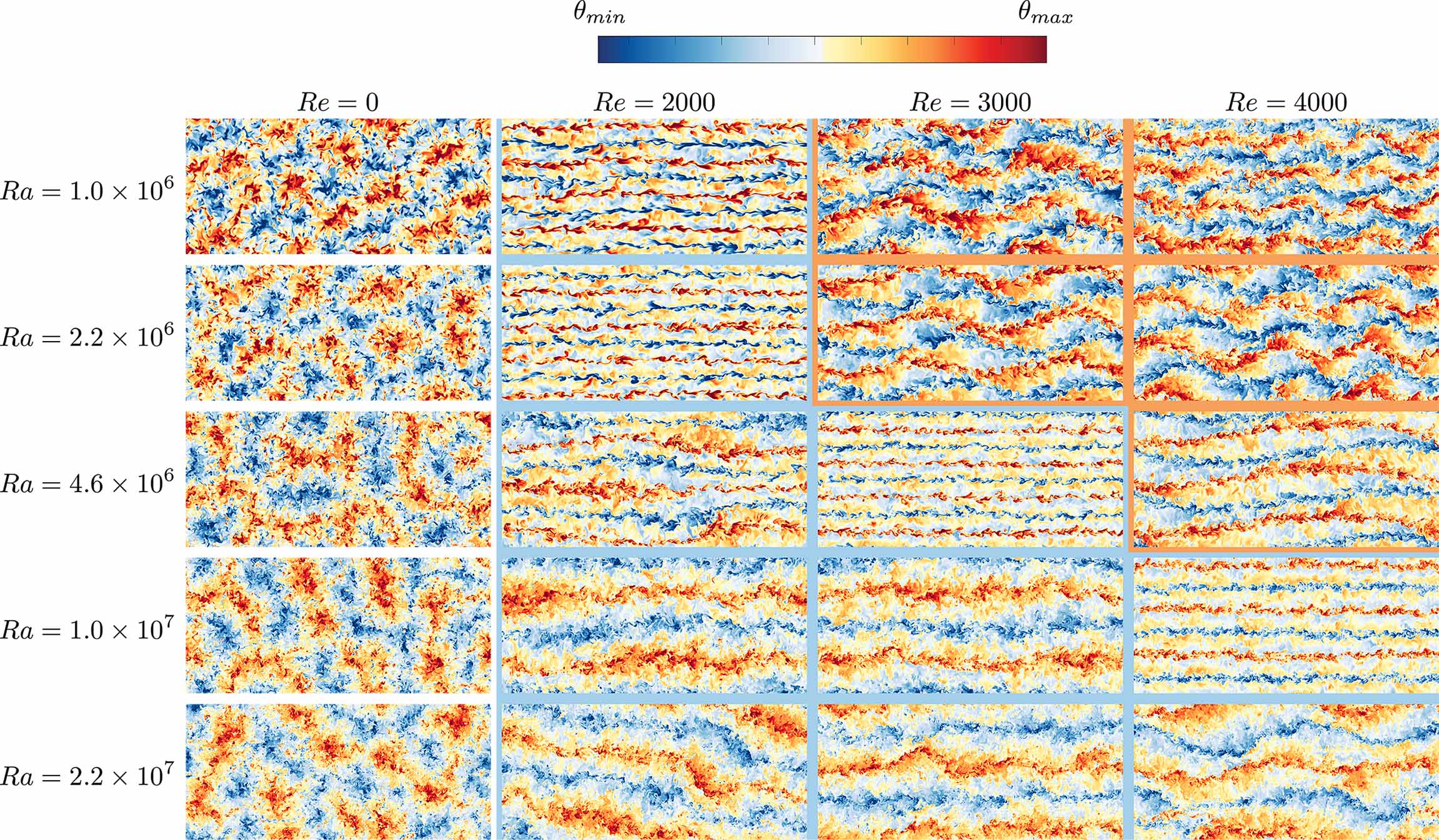}}%
	\caption{\label{fig:temphalfheightoverview} \textcolor{black}{Instantaneous snapshots} of temperature fields at midheight for a subdomain of the parameter space, see figure \ref{fig:status}a and table \ref{tab:gridcases}, focusing on $Ra=1.0\times 10^6 - 2.2\times 10^7$ and $Re_w=0-4000$. The panels have colored borders depending on the flow regime they display: thermal dominated (white), transitional (blue), and shear dominated (orange) regime. \textcolor{black}{For a more detailed quantification of the different flow fields in the presented snapshots, we would like to refer to the values for the Monin-Obukhov length $L_{MO}$ in table \ref{tab:overviewcases}.} An overview of temperature fields over the whole domain can be found in appendix \ref{app:flowfield}. \textcolor{black}{The color ranges of the snapshots in this figure and in figures \ref{fig:ri}, \ref{fig:BLtransition}, \ref{fig:tempenergy}, and \ref{fig:appflowfield} are defined as a variable temperature range for each subfigure to better display the behavior of the thermal structures while keeping the coloring of the structures similar.}}
\end{figure}

\begin{figure}
	\centering
	\includegraphics[width=\textwidth]{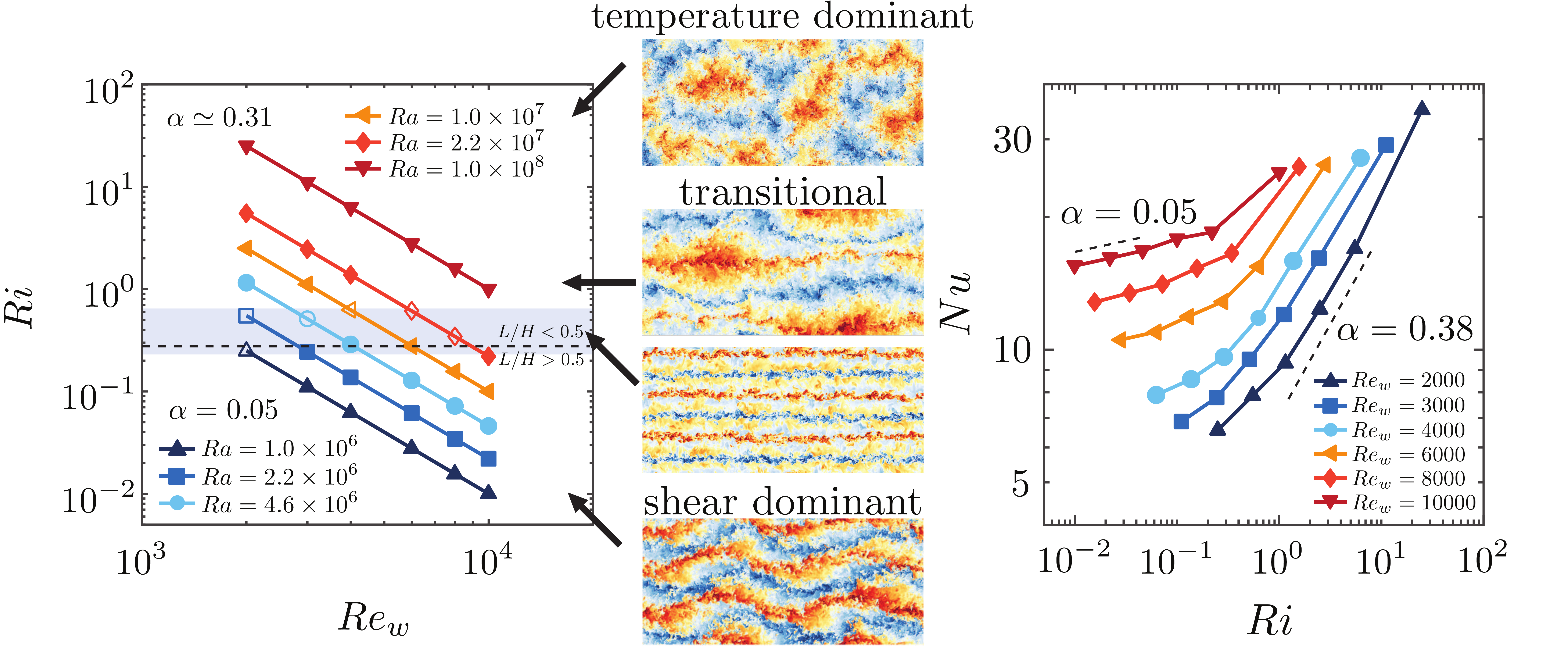}%
	\caption{\label{fig:ri} Left: $Ri$ versus $Re_w$ for different $Ra$. Open symbols indicate the presence of thin straight elongated streaks (see third snapshot from top). The dashed line indicates $L_{MO} =0.5H$. In the middle: \textcolor{black}{instantaneous snapshots of} temperature fields at midheight. Right: $Nu$ as a function of $Ri$ for different $Re_w$. An indication of the effective scaling exponent $\alpha$ in $Nu\sim Ra^\alpha$ in the different regimes is also given. \textcolor{black}{For a more detailed quantification of the different flow regimes in the presented snapshots, we would like to refer to the values for the Monin-Obukhov length $L_{MO}$ in table \ref{tab:overviewcases}.}}
\end{figure}

Following \cite{pir14} we use a von K\'arm\'an constant $\mathcal{K}=0.41$ and $C=5$. The figure shows that the skin friction increases with $Ra$ and decreases with $Re_w$. At fixed $Ra$ the relative strength of the thermal forcing decreases for high $Re_w$ and therefore the obtained friction coefficient converges to the Prandtl law. \textcolor{black}{This agrees very well with \cite{sca15} and \cite{pir17} for buoyant Poiseuille flow.} In figure \ref{fig:cf}b we focus on the data for small $Re_w$. The skin friction in pure Couette flow follows the expected laminar result $C_f=4/Re_w$ \citep{pop00} until a transition to the turbulent state occurs around $Re_w=650-700$. \cite{cer18} discuss that in pipe flow this jump is caused by the formation of puffs and slugs. \cite{bre12} attribute this discontinuous jump in $C_f$ to the lack of restoring forces in plane Couette flow (similar to pipe, channel and boundary layer flows). For the Couette-RB case we do not observe such a discontinuous jump. Instead this sheared RB case is another example, next to the application of Coriolis, buoyancy, and Lorentz forces discussed by \cite{bre12}, which shows that restoring forces can prevent a discontinuous jump in $C_f(Re_w)$. \cite{cha17}, on the other hand, claim that all transitions to turbulence should be continuous if the used box size is large enough. \textcolor{black}{From this figure we can also judge whether a boundary layer is turbulent or not. When the slope of $C_f$ approaches the one of pure Couette flow, the boundary layers are turbulent. Once this slope starts to strongly deviate from the Prandtl law, we consider the boundary layer as not turbulent.}

\begin{figure}
	\vspace{0.5cm}
	\centering
	\includegraphics[height=0.9\textheight]{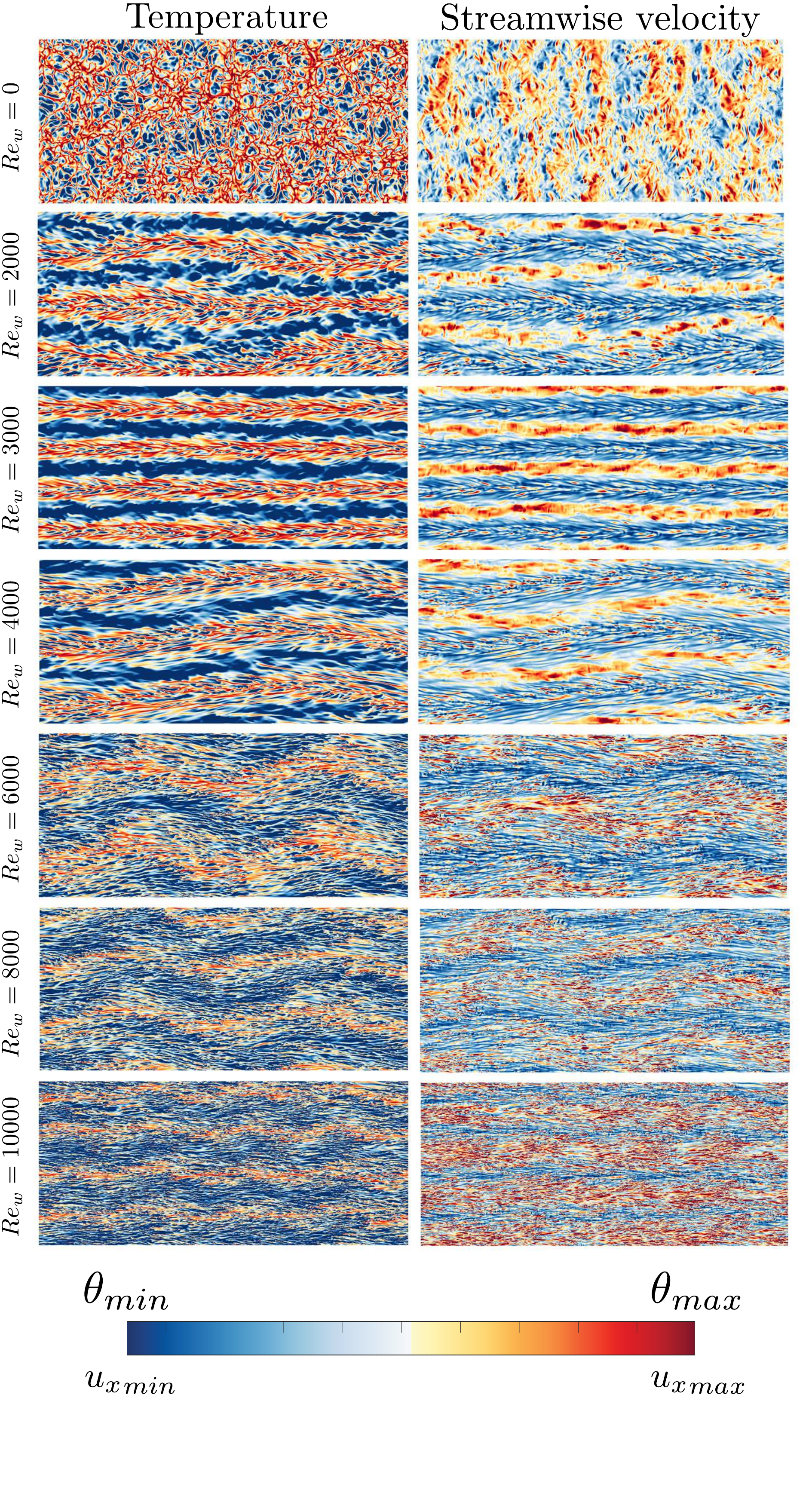}%
	\caption{\label{fig:BLtransition} \textcolor{black}{Instantaneous} near-wall snapshots at $z^+ \approx 0.5$ of the temperature (left) and streamwise velocity (right) for $Ra=4.6 \times 10^6$. $Re_w$ increases from top to bottom. \textcolor{black}{For a more detailed quantification of the different flow fields in the presented snapshots, we would like to refer to the values for the Monin-Obukhov length $L_{MO}$ in table \ref{tab:overviewcases}}.}
\end{figure}

\section{Local flow characteristics} \label{local}

\subsection{Organization of turbulent structures} \label{flowfield}
\noindent To further investigate the dynamics of the different regimes, we show visualizations of the temperature field for all simulations in figure \ref{fig:temphalfheightoverview} and Appendix \ref{app:flowfield}. \textcolor{black}{We chose the mid-height for the location of these two-dimensional snapshots, since this is the position where the flow is the least affected by the wall.} In the thermally dominated regime the primary flow structure resembles the large-scale flow found in RB convection \citep{ste18}. In the transitional regime ($L_{MO} \lesssim 0.5H$), the thermal forcing dominates part of the bulk where large elongated thermal plumes transform into thin straight elongated streaks when $L_{MO}$ approaches $0.5H$. In figure \ref{fig:temphalfheightoverview} and in the appendix this manifests itself as a very visual line diagonally through the diagram, splitting the more thermal- and the more shear dominated cases. In the shear dominated regime ($L_{MO} \gtrsim 0.5H$) we find large-scale meandering structures, similar to the ones found in pressure-driven channel flow with unstable stratification \citep{pir17}. This significant change in flow structure can be linked to the minimum in $Nu$ in figure \ref{fig:nusselt}. The reason for the minimum is that at intermediate shear the thermal convection rolls are broken up, while the shear is not yet strong enough to increase the heat transfer directly. This observation is in agreement with earlier works described above \citep{dom88,sca14,sca15,pir17}.

\textcolor{black}{In figure \ref{fig:ri} we want to present a clear overview over the behavior of the flow structures versus the flow control parameters combined in the bulk Richardson number. On the left side we compare the different values of $Ri$ with the visually observed flow structures. We find a range of $Ri$ in which the flow undergoes a change from the transitional to the shear dominated regime. This happens in a range of $0.2 \lesssim Ri \lesssim 0.7$. In the right panel we can also detect this trend, where the effective scaling of the Nusselt number changes from $Nu \sim Ri^{0.05}$ to $Nu \sim Ri^{0.38}$, but more data points would be necessary to define a more exact point of transition.}

Figure \ref{fig:ri} combines these findings with the above observation that in the shear dominated regime the effective scaling exponent $\alpha$ in $Nu\sim Ra^\alpha$ is much smaller than $1/3$, in the transitional regime $\alpha>1/3$, and in the thermally dominated regime $\alpha \simeq 0.31$. When we compare the regime transitions with the results in figure \ref{fig:nusselt}, it becomes clear that the lowest heat transfer for a given $Ra$ occurs at the end of the transitional regime just before the emergence of the thin straight elongated streaks. Due to the large computational time that is required for each simulation the number of considered cases is limited, which makes it difficult to pinpoint exactly when the heat transfer is minimal and what the flow structure looks like in that case. However, we note that the onset of the shear dominant regime corresponds to the point where the heat transfer starts to increase as the additional shear can then more effectively enhance the overall heat transport.

To get more insight into the boundary layer dynamics in the different regimes, we show the temperature and streamwise velocity at boundary layer height for $Ra=4.6\times 10^6$ in figure \ref{fig:BLtransition}. At this Rayleigh the flow is in the transitional regime for $Re_w=2000$ and $Re_w=3000$, and in the shear dominant regime for $Re_w \geq 4000$. For all cases we observe a clear imprint of the large-scale structures observed at midheight, see figure \ref{fig:temphalfheightoverview} and Appendix \ref{app:flowfield}. This indicates that the large-scale dynamics have a pronounced influence on the flow structures in the boundary layers \citep{ste18}. The figure also reveals that in the transitional and shear dominated regime the lowest temperatures at boundary layer height are observed in the high speed streak regions, which indicates that the regions with the highest shear contribute most to the overall heat flux.

\subsection{Flow statistics} \label{flowstats} 
\begin{figure}
	\centering
	\includegraphics[width=\textwidth]{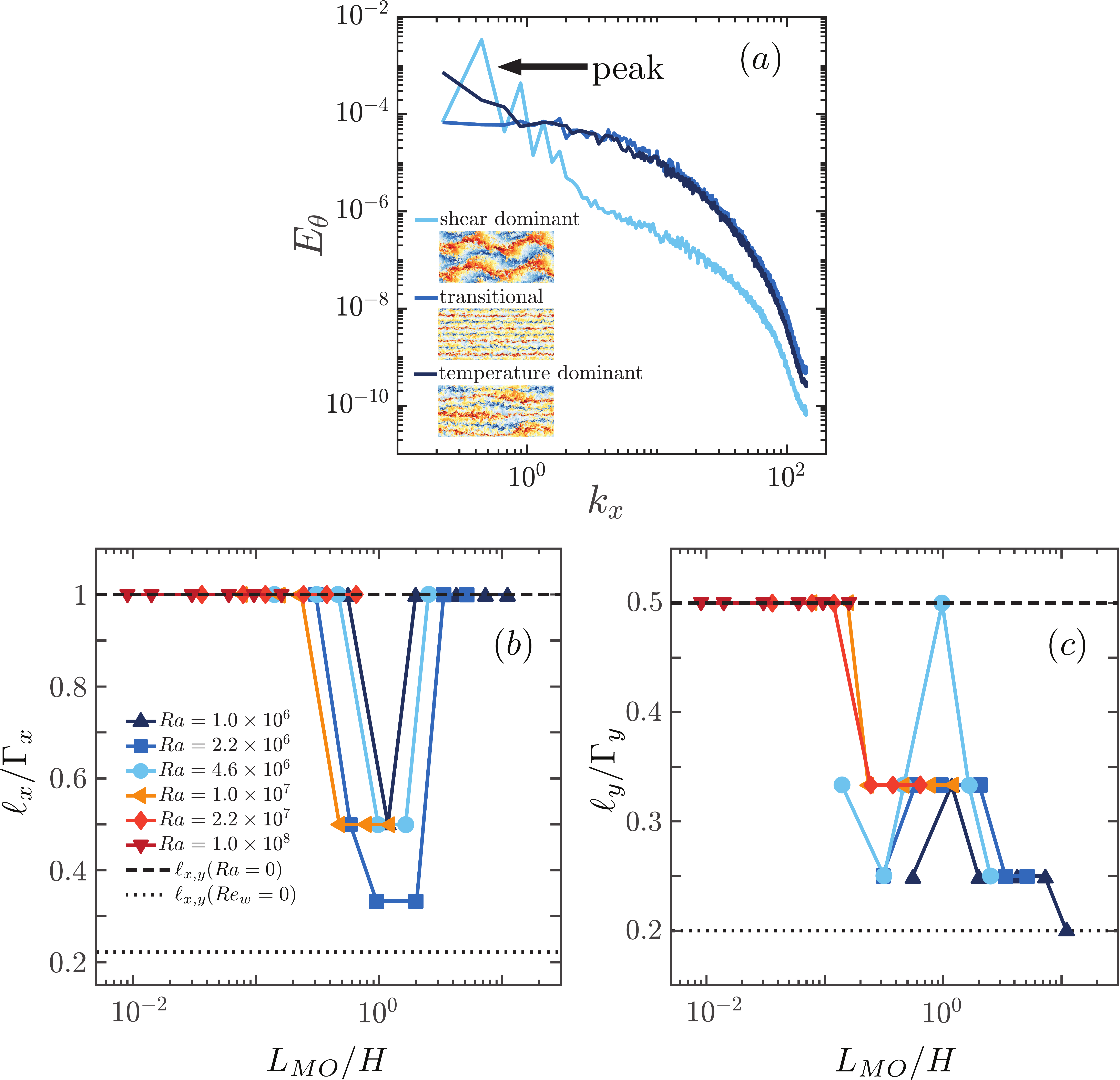}
	\caption{\label{fig:tempenergy} (a) Overview over streamwise temperature variance spectra at $Ra=4.6 \times 10^6$ for different flow regimes \textcolor{black}{at midheight}, (b,c) evolution of peaks in streamwise and spanwise temperature variance spectra, respectively, versus the Monin-Obukhov length. }
\end{figure}

\textcolor{black}{We now present the streamwise temperature variance spectra $E_{\theta} (k)$ in figure \ref{fig:tempenergy} to analyse the size of the large-scale structures as function of the Monin-Obukhov length. The position of the peak in the temperature spectrum indicates the wavelength of the most prominent thermal structure \citep{ste18}. In panel (b) and (c) we plot the evolution of the wavelength of these structures in relation to the absolute size of the flow field. Therefore we define $k_{x_{peak}}$ and $k_{y_{peak}}$ as the wavenumbers of the peak in the respective energy spectrum and $\ell_x =2\pi /k_{x_{peak}}$ and $\ell_y =2\pi /k_{y_{peak}}$ as the respective wavelengths. If the spectrum does not show a clear peak, but keeps growing for small $k$, the structure size is set to the limits of the simulation box, which is $\Gamma_x =9 \pi$ in streamwise (figure \ref{fig:tempenergy}b) and $\Gamma _y=4\pi$ in spanwise direction (figure \ref{fig:tempenergy}c) in this manuscript.}

\textcolor{black}{For $L_{MO} \rightarrow \infty$, $\ell_x \approx \Gamma_x H$, which is expected since for pure Couette flow, structures much larger than $9\pi$ are expected \citep{lee18}. $\ell_y \approx 0.8 \pi H = 0.2 \Gamma_y H$ for the highest shear case, but here more data points are needed for a clearer determination of its behavior. In the other limit of $L_{MO} \rightarrow 0$, i.e. in the transitional regime as the RB case (buoyancy dominated regime) is not shown due to the logarithmic axis, the large-scale structures are elongated over the whole streamwise length, which is consistent with figure \ref{fig:temphalfheightoverview} and figure \ref{fig:appflowfield}. For pure RB convection, where $L_{MO}=0$, $\ell_x$ decreases to $\ell_x \approx 2 \pi H$ and is in agreement with \cite{ste18}. In the spanwise direction, the flow converges already much earlier to the RB case where $\ell_y \approx 2 \pi H = 0.5 \Gamma_y H$.}

\textcolor{black}{In the shear dominated regime, where the flow meanders, the structure size in streamwise direction drops to about half the box length. In spanwise direction this flow regime is present as a local peak in panel (c). Due to the very limited number of datapoints, it is not possible to fully assess the behavior of $\ell_x$ and $\ell_y$ vs $L_{MO}$ for all $Ra$ and $Re_w$. Nevertheless the minimum in $\ell_x$ and peak in $\ell_y$ in the shear dominated regime are very distinct.}

\begin{figure}
	\centering
	\includegraphics[width=\textwidth]{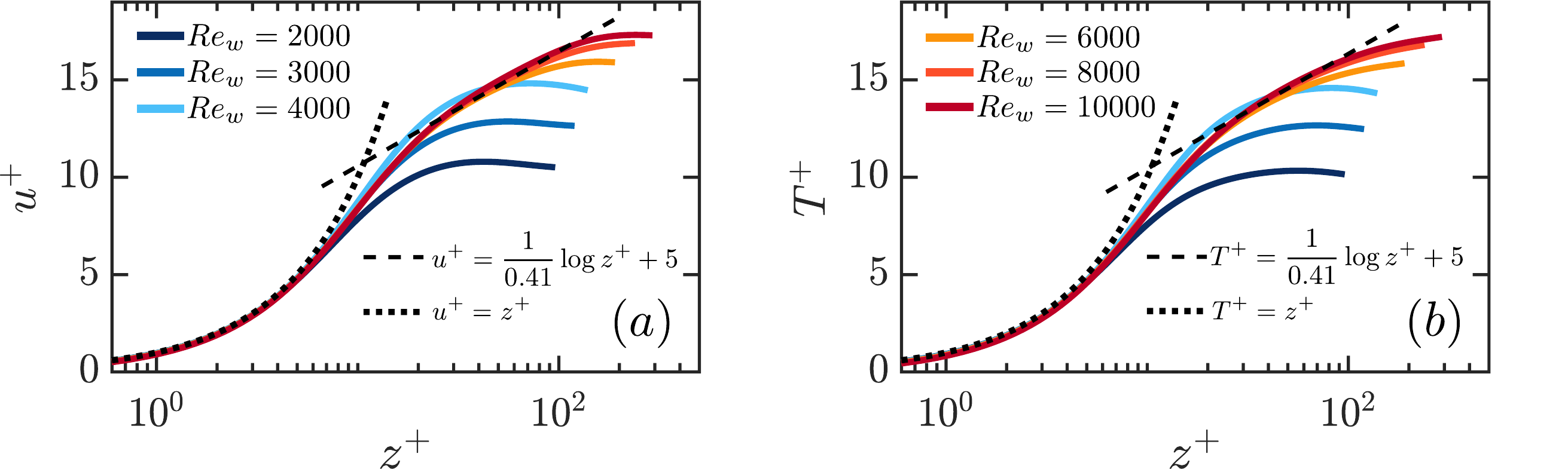}%
	\caption{\label{fig:Ra4e6stats_1} (a) Mean streamwise velocity and (b) temperature profiles, where $u^+ = u/u_\tau$ and, following \cite{pir17}, $T^+ = T/T_\tau$ with the friction temperature $T_\tau = Q/u_\tau$ for $Ra = 4.6 \times 10^6$.}
\end{figure}

To further quantify the cases shown in figure \ref{fig:BLtransition}, we study their flow statistics in figure \ref{fig:Ra4e6stats_1}. It becomes clear that both the temperature and streamwise velocity profiles are not logarithmic in the transitional regime. This indicates that the boundary layers are not turbulent in this state. Hence, the higher $Nu$ scaling in the transitional regime does not seem to be caused by triggering the ultimate regime. In the shear dominated regime the streamwise velocity and temperature profiles seem to converge to a logarithmic profile with increasing $Re_w$ \textcolor{black}{which has also been previously observed in Couette flow \citep{liu03,deb04,cho04,le06} and Poiseuille flow \citep{sca15,pir17} with convection.}

\begin{figure}
	\centering
	\includegraphics[width=\textwidth]{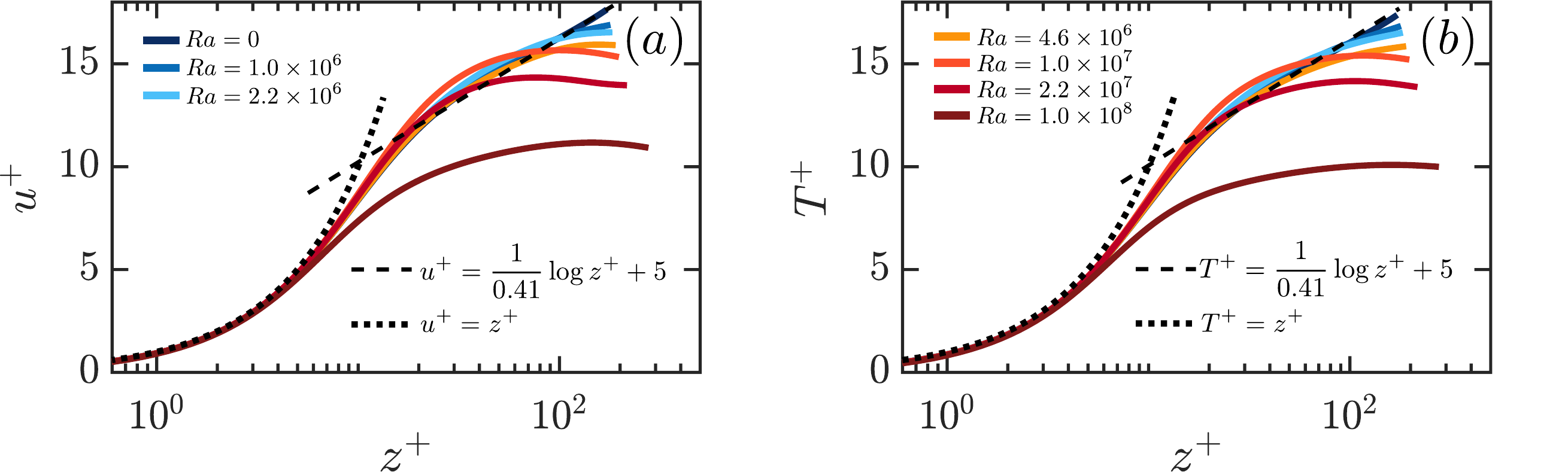}%
	\caption{\label{fig:Re6000stats_1} (a) Mean streamwise velocity and (b) temperature profiles, where $u^+ = u/u_\tau$ and $T^+ = T/T_\tau$ with $T_\tau = Q/u_\tau$ for $Re_w = 6000$. $T^+_{Ra=0}$ was determined through a passive-scalar temperature field.}	
\end{figure}

In figure \ref{fig:Re6000stats_1} we show the same statistical quantities as in figure \ref{fig:Ra4e6stats_1}, but now for fixed $Re_w =6000$. For $Ra \gtrsim 10^7$ the flow is in the transitional regime and for $Ra \lesssim 10^7$ the flow undergoes a transition into the shear dominated regime. Just as in figure \ref{fig:Ra4e6stats_1} we observe that the temperature and streamwise velocity profiles are not logarithmic in the transitional regime. As the Richardson number decreases with decreasing $Ra$, we see that the profiles converge towards a logarithmic behavior. From a comparison with table \ref{tab:overviewcases} we find that $Ri \lesssim 0.2$ seems to be required to achieve logarithmic temperature and velocity profiles. \textcolor{black}{If we refer back to figure \ref{fig:ri}, we can confirm that $Ri \approx 0.2$ is indeed the threshold where the flow undergoes its transition to the shear dominated regime. This is also consistent with the work of \cite{pir17}, who report a regime with increased importance of friction at $Ri \approx 0.1$.} For the parameter regime under investigation the effective scaling exponent $\alpha$ in this regime is well below $1/3$. In both figures we can detect a non-monotonic behavior of both $u^+$ and $T^+$ for low $Re_w$ and high $Ra$. This is connected to an effect of flow layering in the transitional regime where the large-scale thermal plumes get distorted by the shear, but not enough for the meandering structures to evolve and break up this effect. Several further statistical quantities for both constant $Ra$ and constant $Re_w$ have been calculated and can be found in appendix \ref{app:flowstats}.

\section{Concluding remarks} \label{conclusion}
\noindent We performed direct numerical simulations of turbulent thermal convection with Couette type flow shearing. We presented cases in a range $10^6 \leq Ra \leq 10^8$ and $0 \leq Re_w \leq 10^4$, achieving up to $Re_\tau \approx 740$. For fixed Rayleigh number we obtain a non-monotonic progression of $Nu$ similarly to what was previously observed in unstable stratification with a pressure gradient \citep{sca14}. The addition of imposed shear to thermal convection first leads to a reduction of the heat transport by disrupting the turbulent system before the shear becomes strong enough to create meandering streaks that efficiently transport the heat away from the wall. As the impact of the thermal plumes on the flow decreases with increasing shear, the skin friction coefficient at constant $Ra$ drops with increasing $Re_w$.

We find that three flow regimes can be identified in Couette-RB using the Monin-Obukhov length $L_{MO}$ and the thermal boundary layer thickness $\lambda_{\theta}$. In the buoyancy dominated regime ($L_{MO} \lesssim \lambda_{\theta}$) the flow is dominated by large thermal plumes. With decreasing Richardson number we first find a transitional regime ($0.5H \gtrsim L_{MO} \gtrsim \lambda_{\theta}$), before the shear dominated flow regime with large-scale meandering streaks is obtained. For a given $Ra$ the minimum heat transport is found just before the onset of this shear dominated regime when thin straight elongated streaks dominate the flow. We find that in the transitional regime the effective scaling exponent $\alpha$ in $Nu \sim Ra^\alpha$  is larger than $1/3$. An analysis of the flow characteristics shows that the temperature and streamwise velocity profiles are not logarithmic in this transitional regime, which one would expect when this high scaling exponent would indicate the onset of the ultimate regime. Since it is possible to recover logarithmic profiles for low Richardson number flows we want to investigate in future studies whether it is possible to increase the thermal and sheared forcing far enough to trigger ultimate convection in Couette-RB. 

\section*{Acknowledgments}
\noindent We thank Colm-Cille Caulfield and Daniel Chung for fruitful discussions and Jean M. Favre for his support with three-dimensional data visualizations which resulted in figure \ref{fig:3Dpic}. The simulations were supported by a grant from the Swiss National Supercomputing Centre (CSCS) under project ID s713, s802, and s874. This work was financially supported by NWO, by the Dutch center for Multiscale Catalytic Energy Conversion (MCEC), the ERC Advanced Grant ``Diffusive Droplet Dynamics in multicomponent fluid systems", and the Priority Programme SPP 1881 ``Turbulent Superstructures" of the Deutsche Forschungsgemeinschaft. We also acknowledge the Dutch national e-infrastructure SURFsara with the support of SURF cooperative.

\clearpage

\appendix
\section{Flow Field Overview} \label{app:flowfield}
\begin{figure}
	
	\centering
		\rotatebox{90}{\includegraphics[width=0.87\textheight]{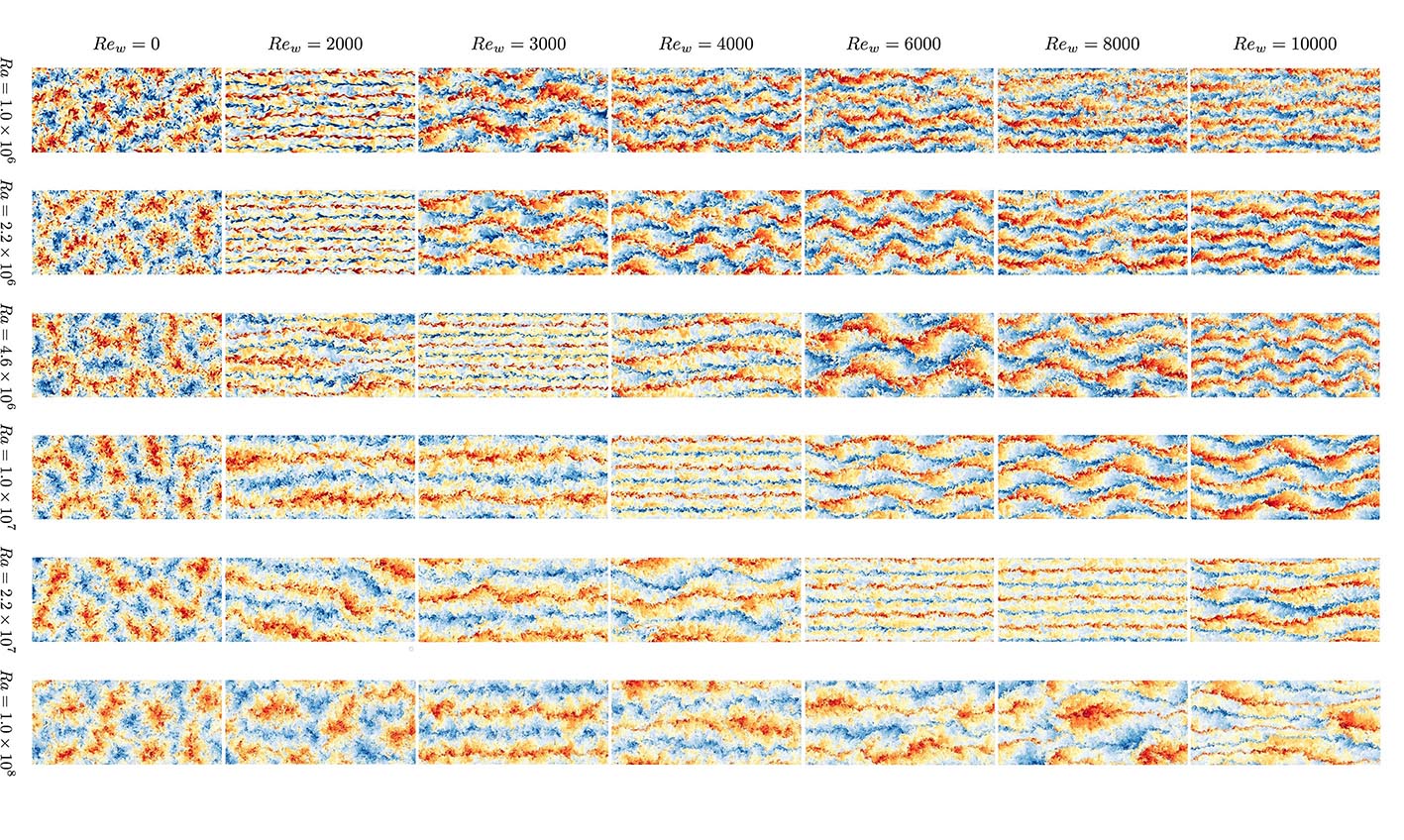}} 
		\caption{\label{fig:appflowfield} \textcolor{black}{Instantaneous snapshots} of all simulated temperature fields at midheight.}		
\end{figure}

\noindent
As an addition to figure \ref{fig:temphalfheightoverview} we present here in figure \ref{fig:appflowfield} the full overview of all temperature fields at midheight, ranging from $Ra=1.0 \times 10^6 - 1.0 \times 10^8$ and $Re_w=0-10000$. All three regimes of thermal domination, transition, and shear domination can be observed here.

\section{Further Flow Statistics} \label{app:flowstats}
\begin{figure}
	\centering
	\includegraphics[width=\textwidth]{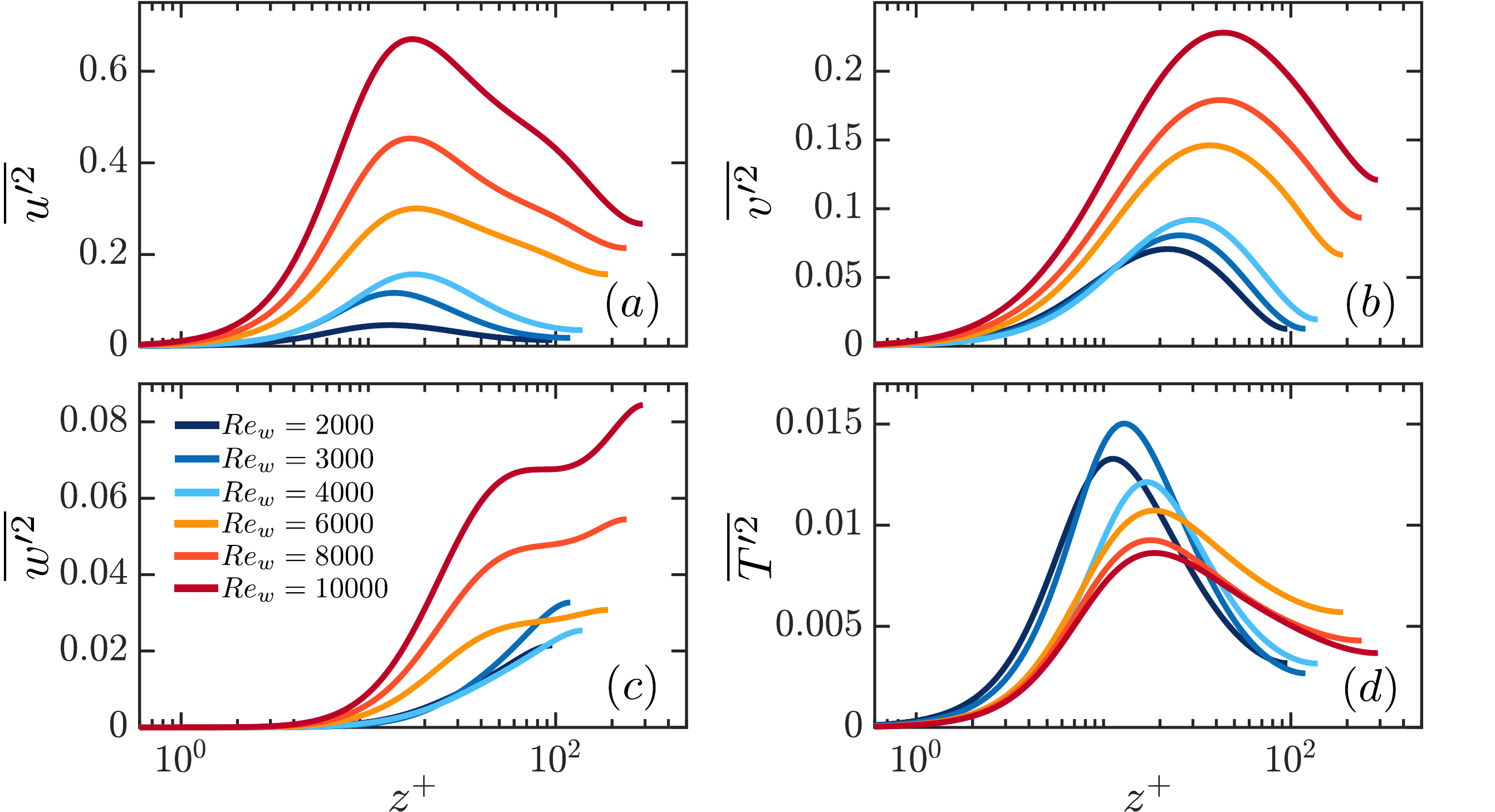}%
	\caption{\label{fig:Ra4e6stats_2} Fluctuations of (a) streamwise, (b) spanwise velocity, (c) wall-normal velocity, and (d) temperature for $Ra=4.6\times 10^6$ in wall units.}
\end{figure}

\noindent
Additionally to figures \ref{fig:Ra4e6stats_1} and \ref{fig:Re6000stats_1} we present further flow statistics in this section. Figure \ref{fig:Ra4e6stats_2} shows the statistical behavior of the flow for constant $Ra=4.6 \times 10^6$ and increasing wall shearing. It can be observed that the velocity fluctuations increase with $Re_w$. The peaks of the temperature fluctuations show a non-monotonic behavior. For low shearing, they first increase with $Re_w$ until it undergoes a transition towards the shear dominated regime, where the temperature fluctuations decrease with increasing wall shearing.

\begin{figure}
	\centering
	\includegraphics[width=\textwidth]{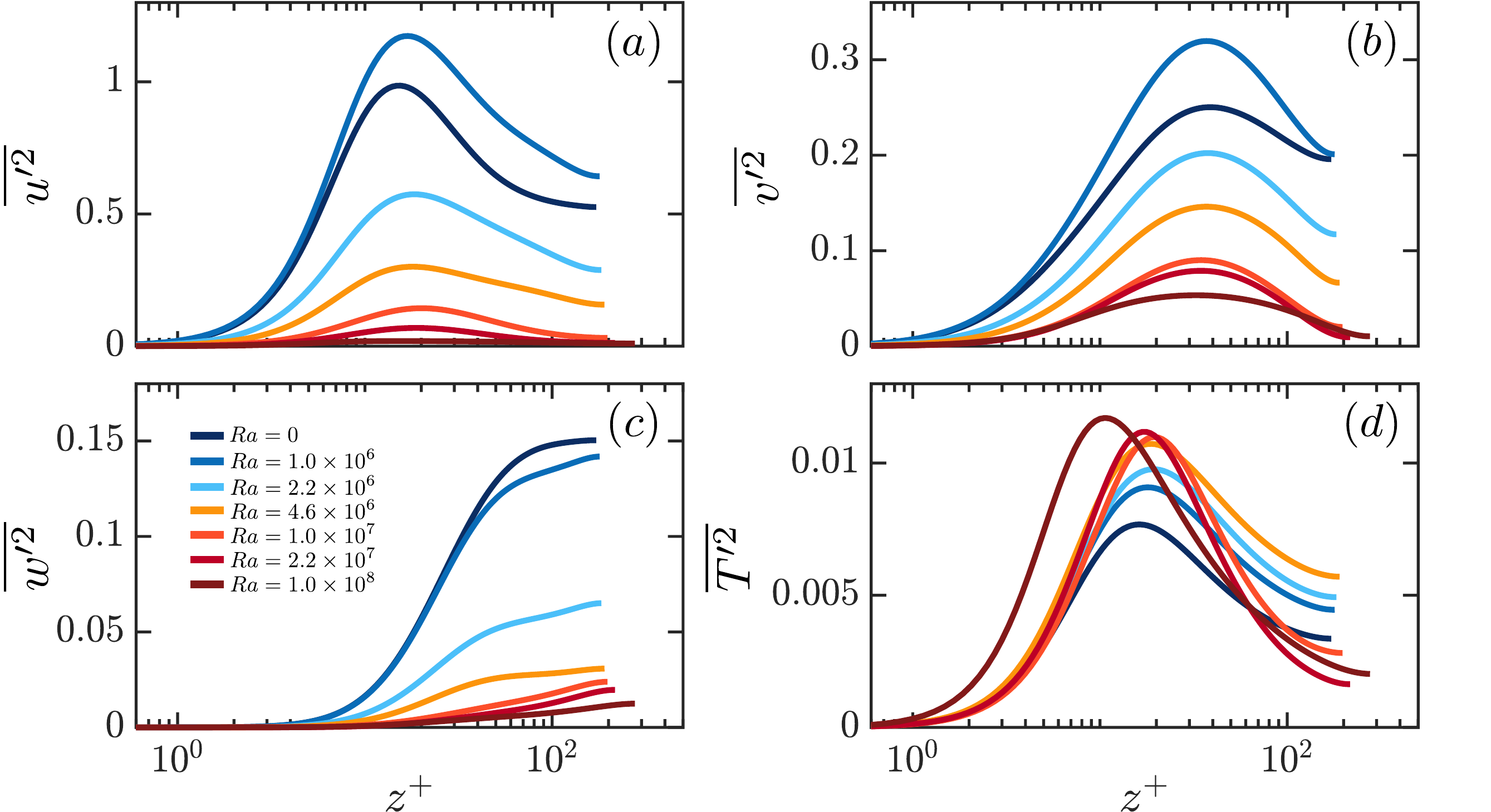}%
	\caption{\label{fig:Re6000stats_2} Fluctuations of (a) streamwise, (b) spanwise velocity, (c) wall-normal velocity, and (d) temperature for $Re_w=6000$ in wall units. $\overline{T'^2}_{Ra=0}$ was determined through a passive-scalar temperature field.}	
\end{figure}

In figure \ref{fig:Re6000stats_2} we present the same flow statistics for constant $Re_w= 6000$ and increasing thermal forcing, starting at plane Couette flow ($Ra=0$). When thermal forcing is added to the Couette flow $u_\textrm{peak}$ and $v_\textrm{peak}$ first increase and then monotonically decrease for increasing $Ra$. Both the wall-normal velocity and the temperature fluctuations decrease completely monotonic for increasing thermal forcing.
  
\section{Monin-Obukhov Fitting} \label{app:Ri_L}
\begin{figure}
	\centering
	\includegraphics[width=0.6\textwidth]{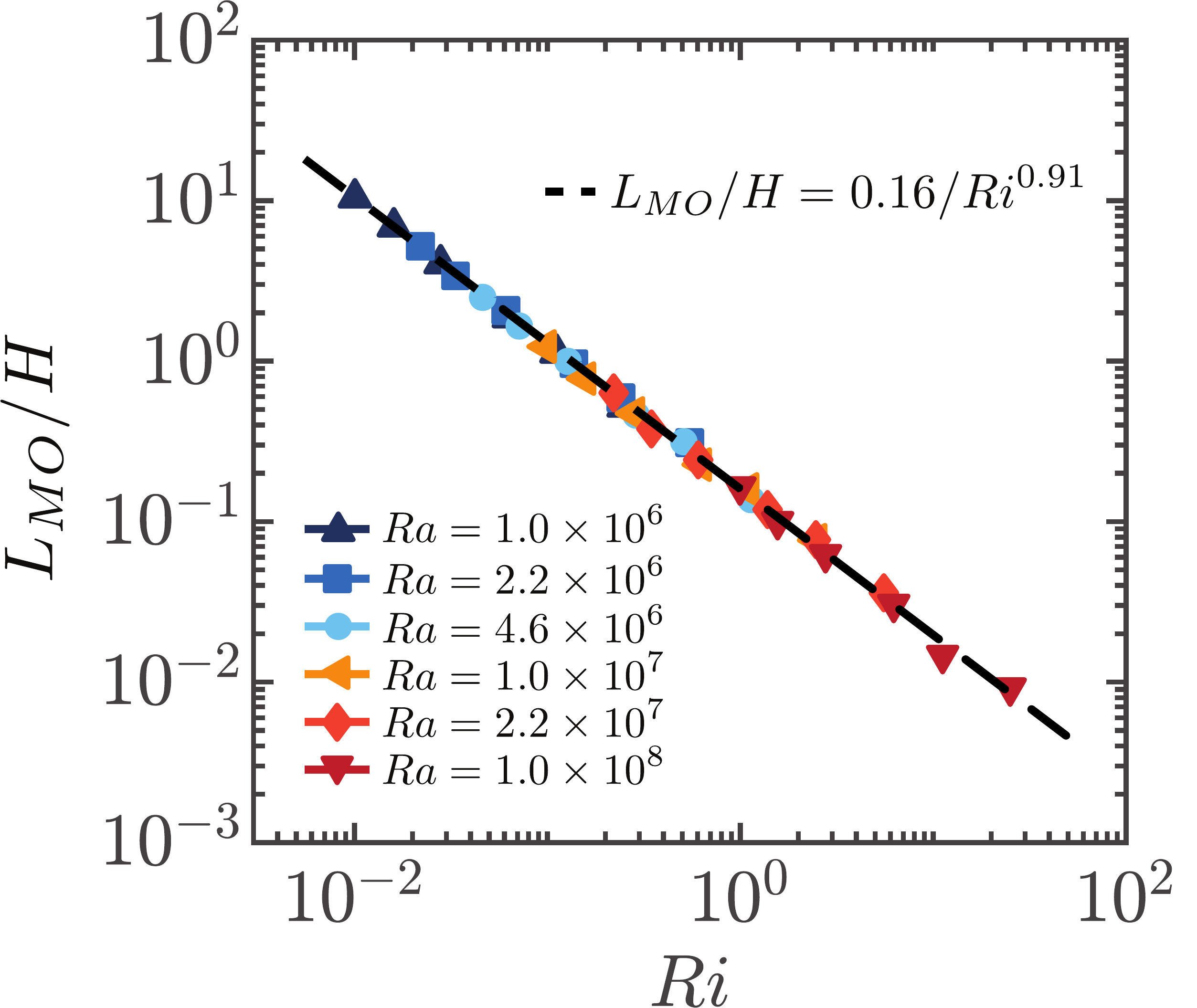}%
	\caption{\label{fig:Ri_L} $L_{MO}/H$ versus $Ri$ of DNS datapoints and the fit $L_{MO}/H=0.16/Ri^{0.91}$.}	
\end{figure}

\noindent
In figure \ref{fig:Ri_L} we present the ratio of shear and thermal forcing in form of the flow output parameter $L_{MO}/H$ versus the flow input parameter $Ri$ from all datapoints of our simulations. We find that the Monin-Obukhov length scales as $L_{MO}/H=0.16/Ri^{0.91}$.

\section{Comparison of $L_{MO}$ and $\lambda_{\theta}$} \label{app:Re_L}
\begin{figure}
	\centering
	\includegraphics[width=0.55\textwidth]{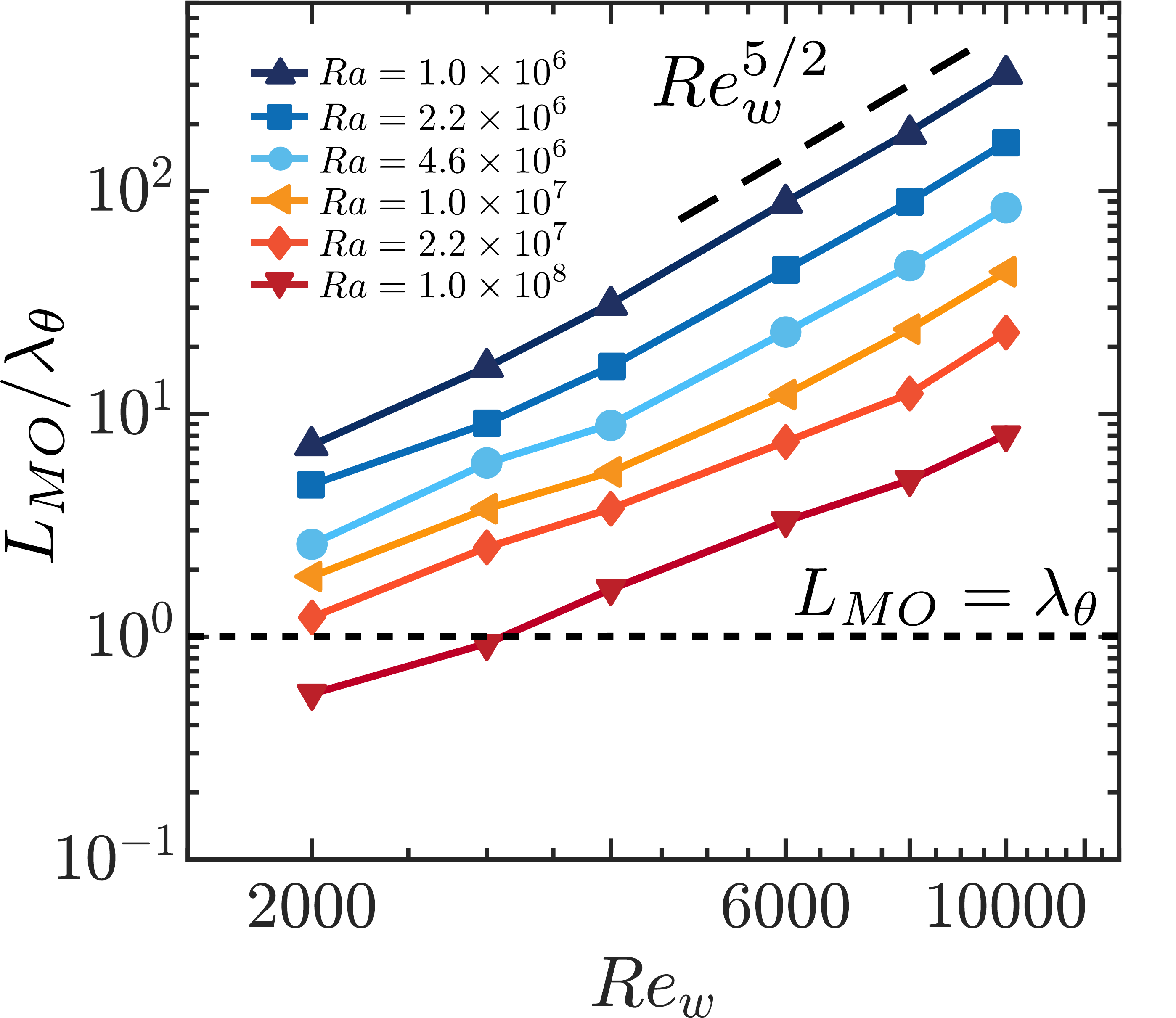}%
	\caption{\label{fig:L_Re} $L_{MO}$ normalized by the thermal boundary layer thickness $\lambda_\theta$ versus $Re_w$. For $L_{MO}/\lambda_{\theta} <1$ the flow is in the thermally dominated regime.}	
\end{figure}

\noindent
In addition to figure \ref{fig:monin}a we present a further visualization of the Monin-Obukhov scale in figure \ref{fig:L_Re}, here normalized by the thermal boundary layer thickness $\lambda_{\theta}$. For $L_{MO}/\lambda_{\theta}<1$ the flow is in the thermally dominated regime. For higher $L_{MO}/\lambda_{\theta}$, the flow first reaches the transitional regime before the shear dominated regime is reached, where $L_{MO}/ \lambda_{\theta} \sim Re_w ^{5/2}$.

\clearpage

\bibliographystyle{jfm}
\bibliography{../../../../../GitHub/Bibliography-turbulence/literature_turbulence}
\end{document}